\documentclass[a4paper,fleqn]{article}
\usepackage[%
  paperwidth=192mm,
  paperheight=262mm,
  vmargin={19mm,19mm},
  hmargin={13.7mm,13.7mm},
  headsep=12pt,
  footskip=12pt,
]{geometry}
\usepackage[authoryear]{natbib}
\usepackage{xcolor}
\usepackage[footnotesize]{caption} 
\usepackage{amssymb}
\usepackage{epsfig}
\usepackage{amsmath}
\usepackage{graphicx}
\usepackage{graphics}
\usepackage{lscape}
\usepackage{float}
\usepackage{subfigure}
\usepackage{multirow}
\usepackage[toc,page]{appendix}
\usepackage{setspace}
\usepackage{color}

\newcommand{\bm}[1]{\mbox{\boldmath $#1$}}
\newcommand{\mc}[1]{\mathcal{#1}}
\newcommand{\la}{\left \langle}
\newcommand{\ra}{\right \rangle}
\newcommand{\be}{\begin{equation}}
\newcommand{\ee}{\end{equation}}
\newcommand{\nn}{{\tt n}}

\begin{document}
\let\WriteBookmarks\relax
\def\floatpagepagefraction{1}
\def\textpagefraction{.001}

\title{A Registration-free approach for Statistical Process Control of 3D scanned objects via FEM}                      

\author{
{\small Xueqi Zhao}\\
{\small Enrique del Castillo\footnote{Corresponding author. Dr. Castillo is Distinguished Professor of Industrial \& Manufacturing Engineering and Professor of Statistics. e-mail: exd13@psu.edu}}\\  
{\small Engineering Statistics and Machine Learning Laboratory}\\
{\small Department of Industrial and Manufacturing Engineering and Dept. of Statistics}\\
{\small The Pennsylvania State University, University Park, PA 16802, USA}}\vspace{0.3cm}

\maketitle

\begin{abstract}
Recent work in on-line Statistical Process Control (SPC) of manufactured 3-dimensional (3-D) objects has been proposed based on the estimation of the spectrum of the Laplace-Beltrami (LB) operator, a differential operator that encodes the geometrical features of a manifold and is widely used in Machine Learning (i.e., Manifold Learning). The resulting spectra are an intrinsic geometrical feature of each part, and thus can be compared between parts avoiding the part to part registration (or ``part localization'') pre-processing or the need for equal size meshes, characteristics which are required in previous approaches for SPC of 3D parts. The recent spectral SPC methods, however, are limited to monitoring surface data from objects such that the scanned meshes have no boundaries, holes or missing portions. In this paper we extend spectral methods  by first considering a more accurate and general estimator of the LB spectrum that is obtained by application of Finite Element Methods (FEM) to the solution of Helmholtz's equation with boundaries. It is shown how the new spectral FEM approach, while it retains the advantages of not requiring part localization/registration or equal size datasets scanned from each part,  it provides more accurate spectrum estimates, which results in faster detection of out of control conditions than earlier methods, can be applied to both mesh or volumetric (solid) scans, and furthermore, it is shown how it can be applied to partial scans that  result in open meshes (surface or volumetric) with boundaries, increasing the practical applicability of the methods. The present work brings SPC methods closer to contemporary research in Computer Graphics and Manifold Learning. MATLAB code that reproduces the examples of this paper is provided in the supplementary materials.

\end{abstract}


\newpage

\section{Introduction}
Modern digital manufacturing deals not only with larger metrology data sets but also more complex data types that have various structures. Point cloud, mesh, and voxel datasets are some of the most common types of data acquired with non-contact sensors. In the area of Quality Control of manufactured parts, methods for the assessment of the quality in a sequence of manufactured parts have relied on point measurements of each part, typically acquired with a coordinate measurement machine (CMM), resulting in point cloud or mesh datasets with the exact same number of points from part to part, so that a point to point correspondence between parts can be established via registration or superposition. While there is considerable work on 3D part inspection, relatively little work has taken place in {\em on-line} inspection or on-line Statistical Process Control (SPC) of manufactured parts based on non-contact sensor data \citep{Babu2017}, and both fields have relied on different types of registration of the parts as a first step or ``pre-processing'' of the scans or measurements obtained from a sequence of parts or between a scanned part and its CAD model \citep{Wells2013,Huang2018, Zang2018}. 

In recent work, \cite{ZhaoEDCTech} (hereafter, ZD) introduce a novel SPC method based on monitoring the spectrum of the Laplace-Beltrami (LB) operator estimated from the scans of each part. As discussed below, the LB operator, present in both the heat  and wave partial differential equations (PDEs) and widely used in Machine Learning (more specifically, in the area of Manifold Learning), codifies the geometrical information of a manifold (in this case, surfaces or solids). The spectrum of the LB operator is intrinsic, that is, it does not depend on the coordinates of the ambient space in which the object is embedded, and hence it can be compared between parts without any registration, totally avoiding the part localization problem. ZD's spectral SPC method is however restricted to meshes modeling the surface of 3-dimensional objects that must be ``closed'', that is, must have no boundaries due to missing parts or holes.
More precisely, we define a {\em closed mesh} as a mesh of discrete elements (flat triangles for surface data and cuboidal voxels for volumetric data) such that all the boundaries of all elements in the mesh are in contact with neighboring elements. Then, we simply define an {\em open mesh} as a mesh that is not closed, hence there are some boundary elements in it. 

In the present paper we extend the spectral SPC method in ZD in three major directions: 1) we estimate the spectrum of the LB operator with Finite Element Methods, resulting in a considerably more accurate estimator of the analytical LB spectrum than the Localized mesh Laplacian used by ZD; 2) we show how the new FEM spectral SPC method can be applied to both surface data (triangulation meshes) and volumetric data (i.e., voxel data) of parts, as acquired by either a range sensor or Computed Tomography (CT) scanner, respectively; 3) we demonstrate how by solving Helmholtz equation as a boundary value problem, the new spectral FEM method can be applied to scans that result in open surface or volumetric meshes. Given that scans are often open due to part regions that are inaccessible to the scanner (a problem which we will refer to as occlusion), the ability to deal with incomplete or open meshes  greatly increases the practical applicability of the proposed spectral methods. The focus therefore is on the more accurate and versatile estimation of the LB spectrum and the consideration of voxel and open meshes; for the specific SPC methods once the spectrum is estimated we follow those used by ZD.

As a preview of the power of the spectral FEM SPC methods presented below to detect defects on manufactured parts relative to earlier methods, consider  the prototype part studied by ZD, shown in Figure \ref{ARL:ClosedPart}, which displays the CAD model followed by three similar parts with different defects, two with small ``chipped'' errors a corner, and one with a small ``protrusion'' in one of the top ``teeth'' of the part. These are triangulation meshes with no boundaries, but as mentioned, the methods presented in this paper apply equally to open triangulation meshes or open volumetric (solid) meshes.  We can compare the power of the different methods  to distinguish between the four types of parts (the non-defective and the three defective) by performing Multidimensional Scaling \citep{borg2005modern} on the first 15 eigenvalues of the LB spectrum computed, for 10 different parts of each type, with the FEM methods presented in this paper and the method used earlier by ZD (based on the localized mesh Laplacian of \cite{li2015localized}). Figure \ref{MDS_parts} is the 3D multidimensional scaling plots of the first 15 LB eigenvalues for the four different part types when different Laplacian discretizations are used. Each part type has 10 simulated realizations, resulting in 10 points of the same color in the abstract 3D Euclidean space. Since multidimensional scaling tries to preserve the between-object distances, this visualization reveals the reason why the FEM methods we propose are more powerful in differentiating subtle shape changes than the \citet{li2015localized} Localized Laplacian, as the leading spectra calculated by the FEM methods are tightly clustered basing on the part type (shown as different colors in the plot), while the LB spectra computed based on the Localized Laplacian is unable to separate the different parts. Hence, if a SPC chart is to monitor the LB spectra as a feature or ``profile'' from each part, it will detect the defective parts considerably faster with the FEM LB methods presented in the present paper.

\begin{figure}
	\centering
		\includegraphics[width=\textwidth]{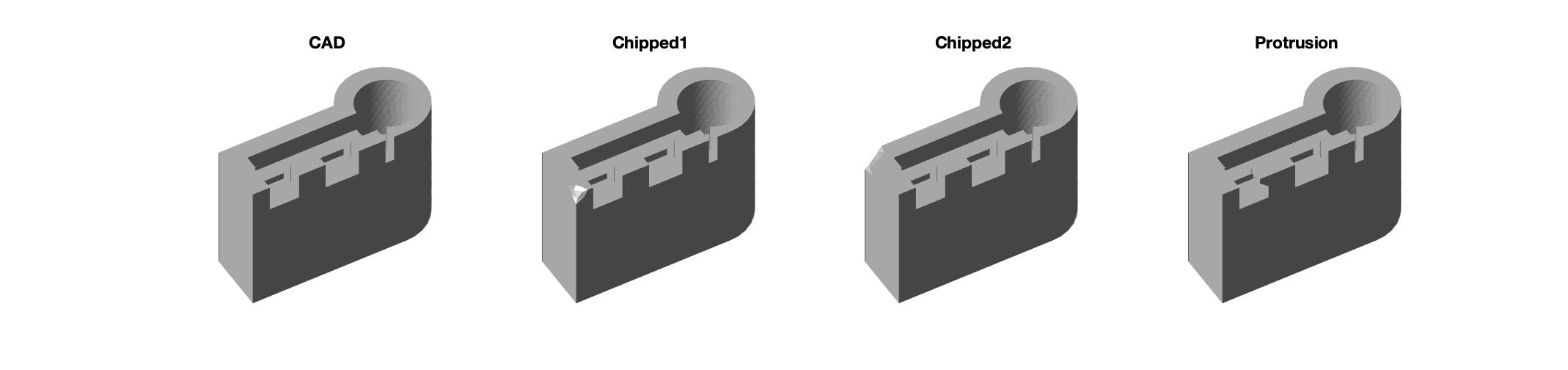}
	\caption{Prototype parts used by \cite{ZhaoEDCTech} to demonstrate their spectral SPC scheme.  The left most part is a perfect part (CAD model), while the other three parts on the right contain different types of defects (``chipped'' corners or a ``protrusion''). All meshes are triangulations without boundaries. This prototype type is used further below for the SPC run length analysis of the new methods in Table \ref{ARL:ClosedPart}.}
	\label{ClosedParts}
\end{figure}

\begin{figure}
	\centering
		\includegraphics[width=\textwidth]{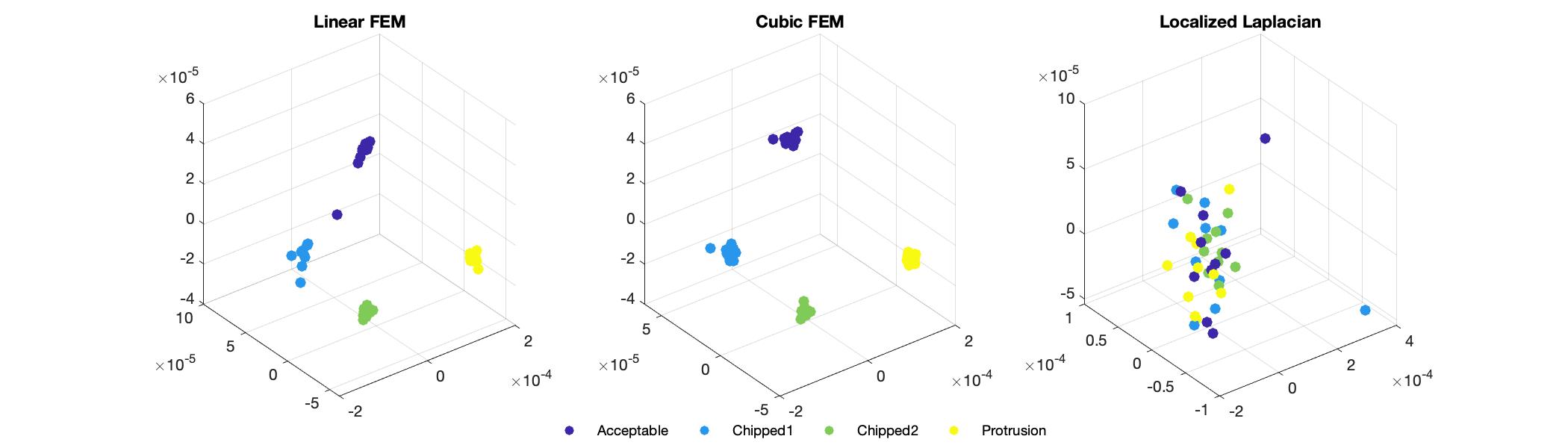}
	\caption{3D multidimensional scaling plot of the first 15 eigenvalues of the LB spectrum for the four part types in Figure \ref{ClosedParts} estimated with the proposed FEM methods in this paper (linear and cubic FEM) and with the Localized Laplacian method due to \cite{li2015localized} used by \cite{ZhaoEDCTech} in their spectral SPC approach. Ten realizations with noise of each part were simulated. The tight clustering of the FEM LB spectra in 4 groups compared to the lack of separation in the spectra of the Localized Laplacian explains why FEM LB-based Statistical Process Control methods can detect defects much faster, while maintaining a low false detection rate. }
	\label{MDS_parts}
\end{figure}

The rest of the paper is organized as follows. Section \ref{sec:2} reviews Differential Geometric notions needed in the sequence, in particular, the definition of the Laplace-Beltrami operator and its properties, which we use in the rest of the paper. Section \ref{sec:3} presents the Finite Element Method (Garlekin approach) for solving a boundary value Helmholtz PDE problem (which contains the LB operator and results from the spatial part of both wave and heat PDEs) and estimates the spectrum of the LB operator in this way. While Finite Element Methods are well known in engineering and science, we present the peculiarities behind both FEM formulations, for surface meshes and voxel data, in enough detail for readers to be able to reproduce our results. Finally, in section \ref{sec:4} we show the performance of an on-line or ``Phase II'' nonparametric SPC chart based on a  sequence of estimated FEM LB spectra of parts, and demonstrate its performance against previous Laplacian methods and SPC methods that are based on registration (superposition) of the parts. As is common in the field of SPC, we are concerned with both quick detection of out of control states (defects) and avoidance of false positives for as long as possible. We close with conclusions and further research in section \ref{sec:5}. The supplementary materials provide MATLAB code that implement our FEM methods and reproduce the examples in the paper.

\section{The Laplace-Beltrami operator and its spectrum}
\label{sec:2}
\subsection{Definition of the LB operator}
To introduce the Laplace-Beltrami operator, consider a  parametric surface ${\bf p}(u,v) = (x(u,v), y(u,v), z(u,v))'$, $(u,v) \in D \subset \mathbb{R}^2$, where $u=x^1$, and $v=x^2$ are local coordinates, defining  $\mc M$, a surface or Riemannian 2-manifold (similar definitions will apply in the case of 3-manifolds or solids). Define the surface differential vectors at ${\bf p}(u,v)$ as:
\[ {\bf p}_u = \frac{\partial {\bf p}(u,v)}{\partial u} = \left(
\frac{\partial x(u,v)}{\partial u},
\frac{\partial y(u,v)}{\partial u},
\frac{\partial z(u,v)}{\partial u}\right)' \quad
\quad \mbox{and} \quad
\quad {\bf p}_v = \frac{\partial {\bf p}(u,v)}{\partial v} = \left(
\frac{\partial x(u,v)}{\partial v},
\frac{\partial y(u,v)}{\partial v},
\frac{\partial z(u,v)}{\partial v}\right)' .\] Define next $
g_{11} = {\la {\bf p}_u, {\bf p}_u\ra} \quad
g_{12} = {\la {\bf p}_u, {\bf p}_v \ra} \quad
g_{22} = {\la {\bf p}_v, {\bf p}_v \ra}$ (where $\la,\ra$ denotes the standard inner product in Euclidean space) and define the Riemannian metric tensor associated with the surface $\mc M$, which defines an inner product on vectors tangent to $\mc M$:
\begin{center}
\begin{tabular}{cc}
$  \la {\bf w}_1, {\bf w}_2 \ra_{\mc{M}} = {\bf w}_1^T {\bm g}  {\bf w}_2 \, \, , \, \,$
&$
{\bm g} = \left(\begin{array}{cc}
g_{11} & g_{12}\\
g_{12} & g_{22}
\end{array} \right)$.\\
\end{tabular}
\end{center}
The metric $\bm g$ is induced by the ambient (Euclidean) space on the surface $\mc M$, but note it is {\em intrinsic}, i.e., it does not rely on the coordinates of the ambient space. Intrinsic geometrical properties, those exclusively based on the metric tensor, are invariant with respect to rigid transformations. Hence, \cite{ZhaoEDCTech}'s idea was to compute an intrinsic differential operator based on scanner data that models geometrical features of an object, because being invariant with respect to rigid transformations it could be used for inspection or statistical quality control without having to register the scanned parts.

The Laplace-Beltrami operator extends the notion of the Laplacian of a function defined on flat (Euclidean) space to functions defined on curved space or Manifolds. Recall the Laplacian of a twice differentiable function $f : \mathbb{R}^n \rightarrow \mathbb{R}$ is minus the divergence of its gradient field: $\Delta f = - \mbox{div}\;  \nabla f = - \sum_{i=1}^n \frac{\partial^2 f}{\partial x_i^2} $ and is evidently a measure of curvature of $f$ at a point $x$. Similarly, for a function $f: \mc M \rightarrow \mathbb{R}$, the {\em  Laplace-Beltrami} (LB) operator is defined as $\Delta_{\mc{M}} f = - \mbox{div}_{\mc{M}} \; \nabla_{\mc{M}} f$, where $\mbox{div}_{\mc{M}} $ is the divergence taken on $\mc{M}$. This is indeed an intrinsic measure of curvature of $f$ defined at a point on the manifold, and in contrast to the Laplacian of a function defined on flat space, it encodes the curvature of the manifold itself as well. In general, applied to a function $f(x^1,...,x^k) \in \mathcal{C}^2$ defined on a $k$-manifold $\mc M$, the LB operator is:
\be
\Delta_{\mc{M}} f = {-}\frac{1}{\sqrt{\mbox{det}(\bm g)}}\sum_{j=1}^k \frac{\partial}{\partial x^j} \left( \sqrt{\mbox{det}(\bm g)} \sum_{i=1}^k g^{ij} \frac{\partial f}{\partial x^i} \right)
\label{LBoperator}
\ee
where $g^{ij}$ are the elements of $\bm g^{-1}$ and det$(\bm g)$ is the determinant of the metric tensor. One important property of the LB operator which aids in its interpretation is that, for a surface (2-manifold):
\begin{equation}
\Delta_\mathcal{M}{\bf p}(u,v) = 2H{\nn}(u,v)
\label{LBnormal}
\end{equation}
and similarly for higher dimensional manifolds, where ${\nn}(u, v)$ is the normal at the point ${\bf p}(u, v)$ on $\mathcal M$ and $H$ is the mean curvature of $\mathcal M$ at $\bf p$, which is the average of the maximum and minimum curvatures in any direction on $\mc M$ from point $\bf p$.

\subsection{The spectrum of the Laplace-Beltrami operator and its use in SPC}
The LB operator appears in both the heat and wave partial differential equations where the space of interest is a Riemannian manifold. In either case, simple separation of variables and consideration only of the spatial variables results in the eigenproblem:
\be
\Delta_{\mc M} f =  \lambda f, \quad 
\label{Helmholtz}
\ee
called the Helmholtz partial differential equation, with an infinite number of eigenfunctions  $f: \mc M \rightarrow \mathbb{R}$ (providing spatial or static solutions to both the heat and wave equations) and corresponding eigenvalues $\lambda \in \mathbb{R}$. We consider solving Helmholtz' equation subject to either Dirichlet boundary conditions ($f = 0$ on the boundary $\Gamma$ of $\mc M$) or Neumann boundary conditions ($\frac{\partial f}{\partial \nn} = \nabla f \cdot \nn = 0$ on $\Gamma$, where $\nn$ is the outward normal vector to $\mc M$ and $\nabla$ is the gradient operator). We point out that considering boundary conditions will permit us to address the case of open or incomplete meshes or volumes, not considered by \cite{ZhaoEDCTech}, whose spectral methods were limited to  closed objects without a boundary.

The collection of eigenvalues $\{\lambda_i\}_{i=0}^{\infty}$ ($0 = \lambda_1 \leq \lambda_2 \leq \lambda_3,.... $) obtained from solving (\ref{Helmholtz}) is called the {\em spectrum} of the LB operator, which \cite{ZhaoEDCTech} proposed to compute numerically to monitor the quality of discrete parts in manufacturing. In the particular case $\mc M \subset \mathbb{R}^2$, the eigenfunctions $f(u,v)$ satisfying the Helmholtz equation are often interpreted as the modes of vibration of a membrane or ``drum'' with resonances at frequencies $\lambda_i$ \citep{Kac}.



Even though the analytical LB spectrum of very few 3D objects is known, one of them being the sphere (see below), the estimated LB spectra can be monitored via a multivariate SPC chart for any object scan (surface or volumetric), comparing the estimated spectrum of a new part produced under regular production (what in SPC is called ``Phase II'') against the spectra estimated from parts obtained while the monitoring scheme was started up (what in SPC is called ``Phase I''). As it will be shown, only the lower part of the spectrum is needed for part to part comparisons. Following the classical SPC paradigm (see, e.g. \cite{MontgomerySPC}) we assume the Phase I spectra were obtained while the process was in a state of statistical control. 

\section{FEM estimation of the LB spectrum}
\label{sec:3}
In contrast with Differential Geometry, we do not have an analytical expression for the object being modeled as a surface (2-manifold) or solid (3-manifold) $\mathcal{M}$, and hence, the first task is to estimate the LB operator from metrology data. In the surface case, we assume in this paper we have available triangulation mesh data (including the possibility of incomplete meshes with holes due to regions on the object that are unreachable to the scanner, see below), consisting of a sample of points from the surface of the object and their adjacency information, usually generated by built-in algorithms used by the sensor mechanism. For solid data, we assume we have available volumetric data acquired by a Computed Tomography (CT) scanner and pre-processed  so the result is a set of 3D set of voxels obtained after application of reconstruction and edge-detection algorithms to find the boundaries of the object from the voxel attenuation values \citep{Kruth2011}.

The analytic LB operator is a differential operator acting on a continuous function. Therefore, the first task in practice is to estimate a {\em discrete} version of the LB operator in the form of a matrix. Several such discretizations exist in the literature, for instance, \citet{delCastillo2020} evaluated the performance of the heat kernel based approximations \citep{belkin2008discrete, li2015localized} for surface data and later suggested to use the Localized LB estimator of \cite{li2015localized} for SPC applications using mesh data,  due to its sparseness \citep{ZhaoEDCTech}. See  \citet{wardetzky2007discrete, patane2016star} for a comprehensive review of discrete LB estimators and their convergence properties.  In this article, we use Finite Element Methods proposed by \citet{reuter2006laplace,reuter2009discrete} who use them in medical applications. FEM methods can be used with either surface or volumetric data \citep{reuter2007global, niethammer2007global}, and, as will be shown below, provide a more accurate estimation of the true analytical LB spectrum and can easily incorporate boundary conditions permitting the estimation of the LB spectrum on open meshes or solids, a key advantage over the \cite{li2015localized} method used in \cite{ZhaoEDCTech}. 

Given the very large literature on Finite Element Methods, we will only present next the details of their application to the specific solution of the Helmholtz equation (\ref{Helmholtz}) using the classical Galerkin variational formulation, from which the spectrum can be obtained, for both surface mesh and volumetric metrology data. Additional details can be found in the Appendix A.

\subsection{Galerkin Variational Formulation for the Helmholtz Boundary Value Problem}

The classical presentation of FEM methods for the solution of a partial differential equation (PDE) starts with the so-called weak, variational, or Galerkin formulation of the problem (see e.g. \cite{Ledret}) which for the Helmholtz equation (\ref{Helmholtz}) we are concerned with in this paper consists in finding a function $f \in V_1$ defined on the manifold $\mc M$ that satisfies the equation:
\begin{equation}
\label{Green_simple}
-\int_\mathcal{M} \nabla f \cdot \nabla \phi\; dV=\lambda\int_\mathcal{M}\phi f\; dV \quad \quad \forall \phi \in V_2
\end{equation}
where $dV$ is either the surface element on a 2-dimensional manifold (surface) or a volume element in a 3-manifold $\mathcal{M}$. 
The weak form (\ref{Green_simple}) is arrived at by considering functions $f$ that satisfy the boundary conditions (see Appendix A for some notes about the derivation of the variational form). The space of functions $V_1$ is called the trial space and $V_2$ is called the test space, and for the Helmholtz equation, they are naturally defined to be both Sobolev spaces $H^1(\mc M) = \{ u: ||\nabla u||^2 + ||u||^2 < \infty \}$ where $||\cdot||$ is the $L_2$ norm. In order to account for the curvature of the manifold, a subtlety about the dot product  $\nabla \phi \cdot \nabla f$ in the first integral, called the first differential parameter of Beltrami (\cite{Kreyszig}, p.230) and sometimes denoted by $\nabla(\phi, f)$,  is that it must be defined as:
$$
\nabla \phi \cdot \nabla f = \partial \phi' \bm{g}^{-1} \partial f = \sum_{i,j} g^{ij}\partial_i\phi\partial_j f
$$
which is an inner product between the gradients of $\phi$ and $f$ each expressed in local basis form (see \cite{Lee2018}, p. 27) that is, $\nabla f = \sum_{i,j} g^{ij} \frac{\partial f}{\partial x^j} {\tt e}_i = \sum_{i,j} g^{ij} \partial_j f { \tt e}_i $ (where $\{{ \tt e}_i\}_{i=1}^d$ is a local Euclidean orthonormal basis)\footnote{The local form of the gradient can be expressed in vector form as ${\bm g}^{-1} \nabla f$ and hence Beltrami's first differential parameter can be written in terms of the inner product defined on $\mc{M}$ as $\la \bm{g}^{-1} \nabla \phi, \bm{g}^{-1} \nabla f \ra_{\mc{M}}$.} and likewise for $\nabla \phi$, a consideration necessary given that gradient vectors are {\em covariant}. 

The customary way to summarize  the weak or variational form is based on defining the inner products:
$$
\la \nabla f, \nabla \phi\ra = \int_{\mc M} \nabla f \cdot \nabla \phi\; dV \quad \mbox{and} \quad \la f, \phi \ra = \int_{\mc M} \phi f\; dV 
$$
so that (\ref{Green_simple}) is usually written in compact form as:
\be
- \la \nabla f, \nabla \phi\ra = \lambda \la f, \phi \ra. 
\label{Variational}
\ee
It can be shown that if $f \in V_1$ is a solution of the Helmholtz equation (\ref{Helmholtz}) then it must satisfy equation (\ref{Variational})  for all $\phi \in V_2$. The reverse implication is also true, it can be shown that if $f \in H_0^1$ satisfies (\ref{Variational}) for all $\phi \in V_2$ then it is a solution of the Helmholtz PDE (\ref{Helmholtz}), see \cite[propositions 4.1 and 4.2 respectively]{Ledret}. Equation (\ref{Green_simple}) can be motivated also as being the Euler-Lagrange equation of either an energy or a least squares error functional, see Appendix A. Solutions $f$ so obtained are eigenfunctions for the Helmholtz equation for the corresponding eigenvalue $\lambda$. 

Rather than directly solving an infinite dimensional problem in functional space, the FEM strategy consists in solving a finite dimensional problem by approximating the solution by $f^N$, a linear combination of $N$ known ``shape'' functions $h_i$, whose coefficients $u_i$ must be determined:
\begin{equation}
\label{form}
f^N = u_1h_1+u_2h_2+\cdots+u_N h_N=\sum_{m=1}^N u_m h_m
\end{equation}
The $N$ linearly independent shape functions $h_1, h_2, \cdots, h_N$ are selected to form a basis for the space of approximate solutions and to be such that each $h_i$ function has local support only over a single discrete finite element in which the space $\mc M$ (surface or volume) is then partitioned.

Substituting (\ref{form}) in the variational problem (\ref{Green_simple}), we choose $N$ different test functions $\phi$ to solve for the $N$ coefficients. In the Garlekin method, the test functions are exactly the same as the shape functions, thus we also substitute $\phi$ for each $h_i (i=1,...,N) $ in (\ref{Green_simple}), which results in the $N$ equations:
\begin{equation}
\begin{aligned}
-\int_\mathcal{M}\sum g^{ij}\partial_i h_l \partial_j \left(\sum_{m=1}^N u_mh_m\right)  dV&=\lambda\int_\mathcal{M}h_l \left(\sum_{m=1}^N u_mh_m\right)  dV, \quad l=1, 2, \cdots, N \\
\sum_{m=1}^N u_m \left(-\int_\mathcal{M}\sum_{i,j} g^{ij}\partial_i h_l\partial_j h_m  dV\right)&=\lambda \sum_{m=1}^N u_m\left(\int_\mathcal{M}h_lh_m dV\right), \quad l=1, 2, \cdots, N
\end{aligned}
\end{equation}
This system of equations can be written as a {\em generalized} eigenvalue problem in the matrix form:
\begin{equation}
\label{AB}
{\bm A \bm U}=\lambda \bm B \bm U
\end{equation}
where $\bm A$ and $\bm B$ are $N$-by-$N$ Gram matrices with entries:
\begin{equation}
\label{ABelements}
A_{lm}=-\la \nabla h_l, \nabla h_m \ra =-\sum_{i,j}g^{ij} \int_\mathcal{M} \partial_i h_l\partial_j h_m dV\quad \mbox{and} \quad 
B_{lm}= \la h_l, h_m \ra
\end{equation}
and $U$ is the vector $(u_1, u_2, \cdots, u_n)^T$. Once the shape functions are chosen, both eigenvalues, which are the Laplace-Beltrami eigenvalues, and eigenvectors, which give the Laplace-Beltrami eigenfunctions, can be easily computed by solving the eigenproblem (\ref{AB}). 
Though $f^N$ is theoretically only an approximation of $f$, given that a finite basis can not span the whole space for infinite dimensional functions, it can become an exact solution when the manifold $\mathcal{M}$ is discretized based on a mesh of finite elements in which case functions on $\mathcal{M}$ are reduced to vectors of dimension $N$, the mesh size. This justifies the choice of the basis size in the previous step.

\subsection{Shape functions used to solve for the LB spectrum--surface case}
Now we discuss how to choose the shape functions when the scan of a part has generated surface data in the form of a triangular mesh. As can be seen from (\ref{form}), the $N$ shape functions compose a basis of the solution space, $\mathbb{R}^N$, with $N$ being the mesh size. The simplest way to ensure linear independency is to use $N$ indicator functions, one for each nodal point. Thus, for $l=1, 2, ..., N$, the $l$th form function, $h_l$, takes value 1 at the $l$th nodal point and 0 at the other points. Its function values elsewhere on $\mathcal{M}$ will be determined later based on the properties of $h_l$. A popular choice for the shape functions are piecewise polynomials, for example:
\begin{itemize}
\item $h_l$ is linear over each finite element
\begin{equation}
\label{linear}
h_l(u,v)=c_{l,1}+c_{l,2}u+c_{l,3}v
\end{equation}
\item $h_l$ is quadratic over each finite element
\begin{equation}
\label{quadratic}
h_l(u,v)=c_{l,1}+c_{l,2}u+c_{l,3}v+c_{l,4}u^2+c_{l,5}uv+c_{l,6}v^2
\end{equation}
\item $h_l$ is cubic over each finite element
\begin{equation}
\label{cubic}
h_l(u,v)=c_{l,1}+c_{l,2}u+c_{l,3}v+c_{l,4}u^2+c_{l,5}uv+c_{l,6}v^2+c_{l,7}u^3+c_{l,8}u^2v+c_{l,9}uv^2+c_{l,10}v^3
\end{equation}
\end{itemize}
where we have used $x^1=u, x^2=v$ as the local coordinates on the manifold $\mathcal{M}$, and the finite elements are simply triangles as we focus on surface triangulations. Note $h_l$ is an indicator function, so for each triangle, $h_l= 1$ only when the triangle has point $l$ as one of its vertices, and $h_l=0$ otherwise. 

\begin{figure}
	\centering
		\includegraphics[width=0.8\textwidth]{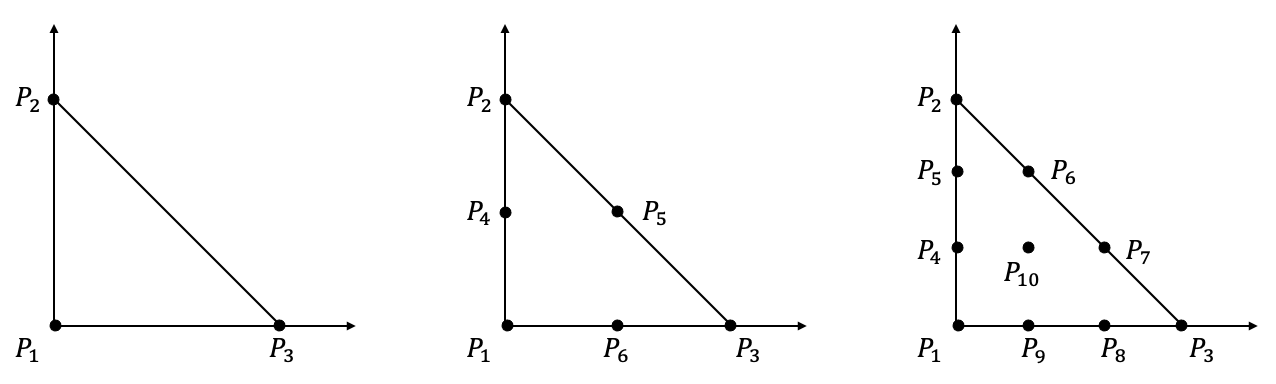}
	\caption{Nodal points used to construct linear shape functions (left), quadratic shape functions (middle), and cubic shape functions (right).}
	\label{nodes}
\end{figure}

Since each triangle is associated with three nodal points at its vertices where the function values $h_l(u,v)$ are known, either 1 or 0 depending on the indices of the shape function under consideration and the vertex, no additional information is needed to uniquely determine the three coefficients $c_{l,j}, j=1,2,3$ in the linear case (\ref{linear}), called the linear FEM method. For the quadratic FEM (\ref{quadratic}), function values at three vertices and at three edge centers will provide the six degrees of freedom needed to uniquely determine the coefficients. For the cubic FEM (\ref{cubic}), since there are ten coefficients, the three vertices, two trisection points on each edge (so six in total), and the triangle centroid are used. The nodal points used in the three different cases are shown in Figure \ref{nodes}. Similar to the linear FEM nodes, each additional node in the quadratic or cubic method corresponds to an additional shape function that takes value 1 at that node and 0 elsewhere. We will focus on the linear FEM and cubic FEM methods in this paper, which are the simplest and the most accurate, respectively.

\subsection{Example: construction of the $\bm A$ and $\bm B$ matrices for surface data (triangulations)}
\label{constructAB}
\begin{figure}
	\centering
		\includegraphics[width=\textwidth]{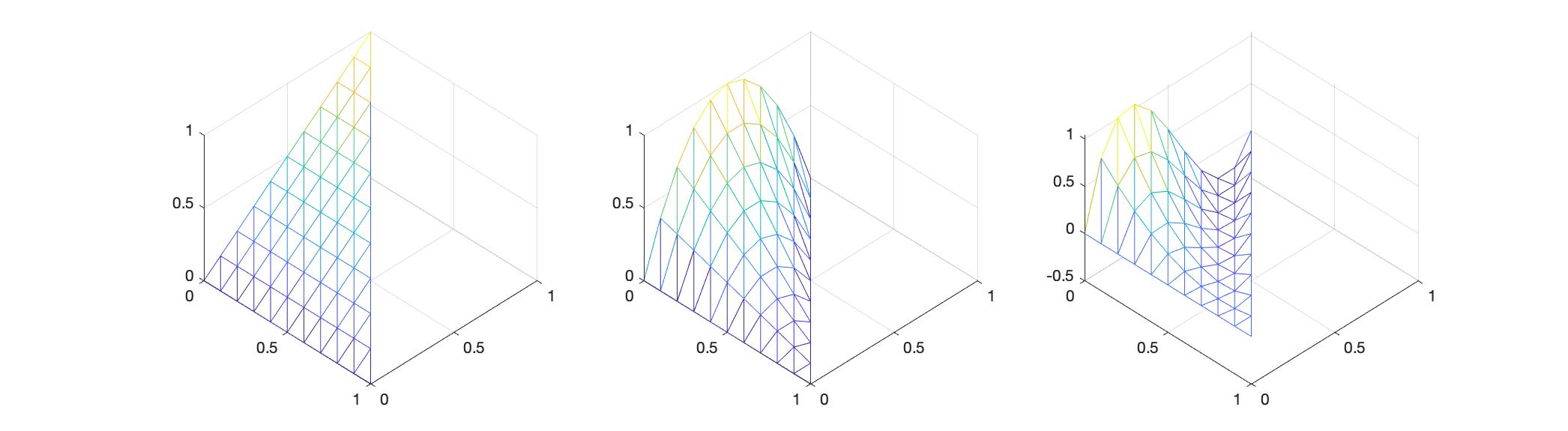}
	\caption{Plots of the shape functions on a unit triangle: $h_2$ in the linear case (left), $h_4$ in the quadratic case (middle), and $h_4$ in the cubic case (right). The z-axis represents function values and lighter colors indicate higher values.}
	\label{form_functions}
\end{figure}

We take the linear FEM as an example to show the construction of the $\bm A$ and $\bm B$ matrices. First let us consider the simplest case, when the mesh only contains one flat triangle as shown in Figure \ref{nodes} (left) and the Euclidean coordinates are $P_1(0,0,0), P_2(0,1,0), P_3(1,0,0)$, respectively. This particular triangle can therefore be described parametrically in local coordinates by:
\begin{equation}
{\bf p}(u,v)=(u,v,0), \qquad 0\leq u\leq 1, 0\leq v\leq 1, u+v\leq 1
\end{equation}
where $(u,v)$ are the surface coordinates which coincide with $(x,y)$ in this case. Using this local parametrization, $P_1={\bf p}(0,0), P_2={\bf p}(0,1), P_3={\bf p}(1,0)$, and the metric tensor of this plane is
\begin{equation}
\begin{cases}
\frac{\partial {\bf p}(u,v)}{\partial u}=(1,0,0) \\
\frac{\partial {\bf p}(u,v)}{\partial v}=(0,1,0)
\end{cases}
\quad\Rightarrow\quad
\begin{cases}
g_{11}=\frac{\partial {\bf p}(u,v)}{\partial u}\cdot\frac{\partial {\bf p}(u,v)}{\partial u}=1\\
g_{12}=\frac{\partial {\bf p}(u,v)}{\partial u}\cdot\frac{\partial {\bf p}(u,v)}{\partial v}=0\\
g_{22}=\frac{\partial {\bf p}(u,v)}{\partial v}\cdot\frac{\partial {\bf p}(u,v)}{\partial v}=1
\end{cases}
\quad\Rightarrow\quad
{\bf g}=\begin{pmatrix} 1 & 0 \\ 0 & 1\end{pmatrix}=I
\end{equation}
which is not surprising since the triangle is flat. Since there are three points, we need three shape functions $h_1$, $h_2$, and $h_3$. Taking the linear case, $h_1(u,v)=c_{1,1}+c_{1,2}u+c_{1,3}v$ as an example, which takes value 1 only at $P_1$ and 0 elsewhere, we obtain the following system of equations
\begin{equation}
\label{h1}
\begin{cases}
h_1(P_1)=h_1(0,0)=c_{1,1}=1\\
h_1(P_2)=h_1(0,1)=c_{1,1}+c_{1,3}=0\\
h_1(P_3)=h_1(1,0)=c_{1,1}+c_{1,2}=0\\
\end{cases} \quad\Rightarrow\quad
\begin{cases}
c_{1,1}=1\\
c_{1,3}=-1\\
c_{1,2}=-1\\
\end{cases} \quad\Rightarrow\quad
h_1(u,v) = 1-u-v 
\end{equation}
Similarly, we can solve for $h_2$ and $h_3$:
\begin{equation}
\label{h23}
h_2(u,v)=v, \qquad h_3(u,v)=u
\end{equation}
Figure \ref{form_functions} plots some of the shape functions obtained in this way. As the degree of the polynomials increases, the shape functions become more flexible, as expected. Once the analytical expressions for the shape functions are known, entries in matrices $\bm A$ and $\bm B$ can be easily calculated. 

Consider now the case of three points with nontrivial coordinates: $P_1(x_1,y_1,z_1), P_2(x_2,y_2,z_2), P_3(x_3,y_3,z_3)$. This can be converted to the previous case by modifying only the surface parametrization:
\begin{equation}
\begin{aligned}
{\bf p}(u,v)=\left(x_1+u(x_3-x_1)+v(x_2-x_1),y_1+u(y_3-y_1)+v(y_2-y_1),z_1+u(z_3-z_1)+v(z_2-z_1)\right), \\
0\leq u\leq 1, 0\leq v\leq 1, u+v\leq 1
\end{aligned}
\end{equation}
with which we still have $P_1={\bf p}(0,0)$, $P_2={\bf p}(0,1)$, and $P_3={\bf p}(1,0)$. Consequently, the analytical expressions for linear $h_1$, $h_2$, and $h_3$ remain the same as in (\ref{h1}) and (\ref{h23}), so there is no need to reevaluate the integrals in (\ref{ABelements}). On the other hand, the metric tensor $\bf g$ changes with the parametrization:
\begin{equation}
\label{tensor}
\begin{cases}
\frac{\partial {\bf p}(u,v)}{\partial u}=(x_3-x_1,y_3-y_1,z_3-z_1) \\
\frac{\partial {\bf p}(u,v)}{\partial v}=(x_2-x_1,y_2-y_1,z_2-z_1)
\end{cases}
\quad\Rightarrow\quad
{\bf g}=\begin{pmatrix} \|P_3-P_1\|^2 & (P_3-P_1)\cdot(P_2-P_1) \\ (P_3-P_1)\cdot(P_2-P_1) & \|P_2-P_1\|^2\end{pmatrix}
\end{equation}
This affects the entries of the $\bm A$ and $\bm B$ matrices in two ways. First, obviously $g^{ij}$ differs from triangle to triangle. Secondly, the surface area element in the integrals changes too, since $dV=\sqrt{\det({\bf g})}dudv$. So eq (\ref{ABelements}) becomes
\begin{equation}
\small
\begin{aligned}
\label{ABelements_general}
A_{lm}&=-\int_\mathcal{M}\sum_{i,j} g^{ij}\partial_i h_l\partial_j h_m\sqrt{\det({\bf g})}dudv\\
&=-\sqrt{\det({\bf g})}\left(g^{11}\int_\mathcal{M}\partial_uh_l \partial_uh_mdudv+g^{12}\int_\mathcal{M}\partial_u h_l\partial_vh_mdudv+g^{21}\int_\mathcal{M}\partial_vh_l\partial_uh_mdudv+g^{22}\int_\mathcal{M}\partial_vh_l\partial_vh_mdudv\right)\\
B_{lm}&=\sqrt{\det({\bf g})} \int_\mathcal{M}h_lh_m dudv.
\end{aligned}
\end{equation}
Finally we consider the case of meshes consisting of an arbitrary number of connected triangles. Note that the $(l,m)$th elements in matrices $\bm A$ and $\bm B$ require $\partial_i h_l\partial_j h_m$, and $h_lh_m$ respectively, which are nonzero only when points $l$ and $m$ are connected by an edge and therefore appear in the same triangle(s). This implies that we can process the mesh triangle by triangle, and only fill in the entries of $\bm A$ and $\bm B$ as needed. Furthermore, as we discussed above, for each triangle, only the metric tensor $\bf g$ (\ref{tensor}) needs to be recalculated. When an edge connecting two points, say $l$ and $m$, is not on the boundary, it will be included in two adjacent triangles, each of which gives a value for $a_{lm}$ and $b_{lm}$. In this case, the two different values of $a_{lm}$ (or $b_{lm}$) are the integrals of $\partial_i h_l\partial_j h_m$ (or $h_lh_m$) evaluated in the two individual triangles, respectively, and thus should be added up.

\subsection{Shape functions for volumetric (voxel) data}
\begin{figure}
	\centering
		\includegraphics[width=\textwidth]{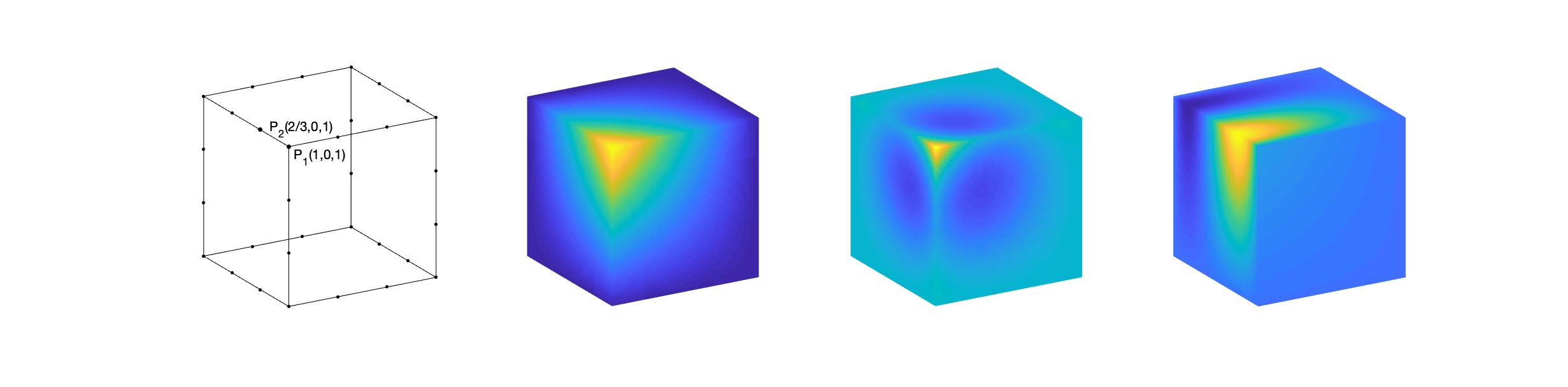}
	\caption{Nodal points and plots of the shape functions on a voxel. From left to right: a) Nodal points with two points labeled in their Euclidean coordinates, $P_1(1,0,1)$ and $P_2(2/3,0,1)$. The 8 vertices are used for linear functions while all 32 nodal points shown are used for cubic functions. b) Linear $h_1$ corresponding to $P_1$. c) Cubic $h_1$ corresponding to $P_1$. d) Cubic $h_2$ corresponding to $P_2$. Lighter colors indicates higher function values. The graphs show how shape functions ensure $h_l(P_m)=0$ for $l\neq m$.}
	\label{FormVol}
\end{figure}

In addition to their use for computing the LB spectrum for 2-dimensional surface data, the FEM methods can be easily extended to the voxel or volumetric data case. The shape functions have three variables now with the increased dimension, and as suggested by \citet{reuter2007global} we use the trilinear function and the cubic function of the serendipity family \citep{arnold2011serendipity} for the linear FEM and cubic FEM, respectively:
\begin{equation}
\begin{aligned}
h_l(u,v,w)=&c_{l,1}+c_{l,2}u+c_{l,3}v+c_{l,4}w+c_{l,5}uv+c_{l,6}uw+c_{l,7}vw+c_{l,8}uvw \\
h_l(u,v,w)=&c_{l,1}+c_{l,2}u+c_{l,3}v+c_{l,4}w+c_{l,5}uv+c_{l,6}uw+c_{l,7}vw+c_{l,8}uvw\\
&+c_{l,9}u^2+c_{l,10}u^2v+c_{l,11}u^2w+c_{l,12}u^2vw+c_{l,13}v^2+c_{l,14}v^2u+c_{l,15}v^2w+c_{l,16}v^2uw\\
&+c_{l,17}w^2+c_{l,18}w^2u+c_{l,19}w^2v+c_{l,20}w^2uv+c_{l,21}u^3+c_{l,22}u^3v+c_{l,23}u^3w+c_{l,24}u^3vw\\
&+c_{l,25}v^3+c_{l,26}v^3u+c_{l,27}v^3w+c_{l,28}v^3uw+c_{l,29}w^3+c_{l,30}w^3u+c_{l,31}w^3v+c_{l,32}w^3uv
\end{aligned}
\end{equation}
Similar to the mesh case, a linear shape function uses the 8 vertices of each finite element, a voxel in this case, to uniquely determine its 8 coefficients, while a cubic shape function needs 24 more nodal points, 2 trisection nodes on each of the 12 edges, together with the original 8 vertices to provide a total of 32 degrees of freedom (first graph in Figure \ref{FormVol}). Each shape function is the indicator function of a corresponding node, and each is again a piecewise polynomial as before. Figure \ref{FormVol} shows a plot of fitted linear and cubic shape functions in a voxel, where lighter colors indicate higher function values. Since the voxel and the nodal points we use are highly symmetric, the linear shape functions have only one pattern, as in the second graph, while the cubic shape functions have two patterns, depending on whether the node is on a corner or an edge, both of which are shown in the last two graphs. Note the color scales differ in different graphs for a more detailed illustration of how function values vary within each object, but it can inferred that whichever color that occurs on the edges corresponds to a function value of 0 on that particular graph.

The entries of the metric tensor $\bf g$ for voxels are simpler than for triangles. A voxel is essentially a cuboid of fixed dimensions and can be parametrized by:
\begin{equation}
\begin{aligned}
{\bf p}(u,v,w)=\left(x+s_1 u, y+s_2 v, z+s_3 w\right), 0\leq u\leq 1, 0\leq v\leq 1, 0\leq w\leq 1
\end{aligned}
\end{equation}
where $(x,y,z)$ are the coordinates of the cuboid vertex that is closest to the origin, and $s_1$, $s_2$, and $s_3$ are the three edge lengths of the cuboid, respectively. Then the metric tensor is 
\begin{equation}
\begin{cases}
\frac{\partial {\bf p}(u,v,w)}{\partial u}=(s_1,0,0) \\
\frac{\partial {\bf p}(u,v,w)}{\partial v}=(0,s_2,0) \\
\frac{\partial {\bf p}(u,v,w)}{\partial w}=(0,0,s_3) \\
\end{cases}
\quad\Rightarrow\quad
\begin{cases}
g_{11}=\frac{\partial {\bf p}(u,v,w)}{\partial u}\cdot\frac{\partial {\bf p}(u,v,w)}{\partial u}=s_1^2\\
g_{12}=\frac{\partial {\bf p}(u,v,w)}{\partial u}\cdot\frac{\partial {\bf p}(u,v,w)}{\partial v}=0\\
g_{13}=\frac{\partial {\bf p}(u,v,w)}{\partial u}\cdot\frac{\partial {\bf p}(u,v,w)}{\partial w}=0\\
g_{22}=\frac{\partial {\bf p}(u,v,w)}{\partial v}\cdot\frac{\partial {\bf p}(u,v,w)}{\partial v}=s_2^2\\
g_{23}=\frac{\partial {\bf p}(u,v,w)}{\partial v}\cdot\frac{\partial {\bf p}(u,v,w)}{\partial w}=0\\
g_{33}=\frac{\partial {\bf p}(u,v,w)}{\partial w}\cdot\frac{\partial {\bf p}(u,v,w)}{\partial w}=s_3^2\\
\end{cases}
\quad\Rightarrow\quad
{\bf g}=\begin{pmatrix} s_1^2 & 0  & 0 \\ 0 & s_2^2 & 0 \\ 0 & 0 & s_3^2 \end{pmatrix}
\end{equation}
Note ${\bf g}$ is independent of $(x,y,z)$, that is, independent of the location of the voxel, and depends only on the size of the voxel, which is constant. Therefore, we can calculate only once the matrix elements $a_{lm}$ and $b_{lm}$ in a voxel and fill in all the other entries in $\bm A$ and $\bm B$ computing (\ref{ABelements_general}) by simply looking up the global nodal point indices (each nodal point has an index local within each voxel and a global index within the whole volume of voxels).


\subsection{Properties of the FEM methods}
\subsubsection{Sparsity and symmetry}
\label{SparsitySymmetry}
As can be seen in (\ref{ABelements}), $b_{lm}$ is nonzero if and only if both $h_l$ and $h_m$ are not constantly zero over at least one finite element (a triangle or a voxel), which happens only when point $l$ and $m$ appear in the same finite element. This indicates that matrix $B$ is sparse. For example, in the surface linear FEM method, the number of nonzero elements along row $l$ or column $l$ equals the degree, or number of neighbors, of point $l$. Similarly, $a_{lm}$ is nonzero if and only if both $h_l$ and $h_m$ are not constant over at least one finite element. Since $h_l$ is an indicator function and piecewise polynomial over each finite element, it is zero for elements not associated with point $l$, and it is never a constant function for elements associated with point $l$, $l=1, 2, \cdots, N$. Therefore, when $h_l$ is non-constant (has a gradient) it is when it is nonzero, thus $a_{lm}$ and $b_{lm}$ are zero or nonzero at the same time. In other words, matrix $\bm A$ is sparse as well and has the same nonzero structure as matrix $\bm B$.

\begin{figure}
	\centering
		\includegraphics[width=\textwidth]{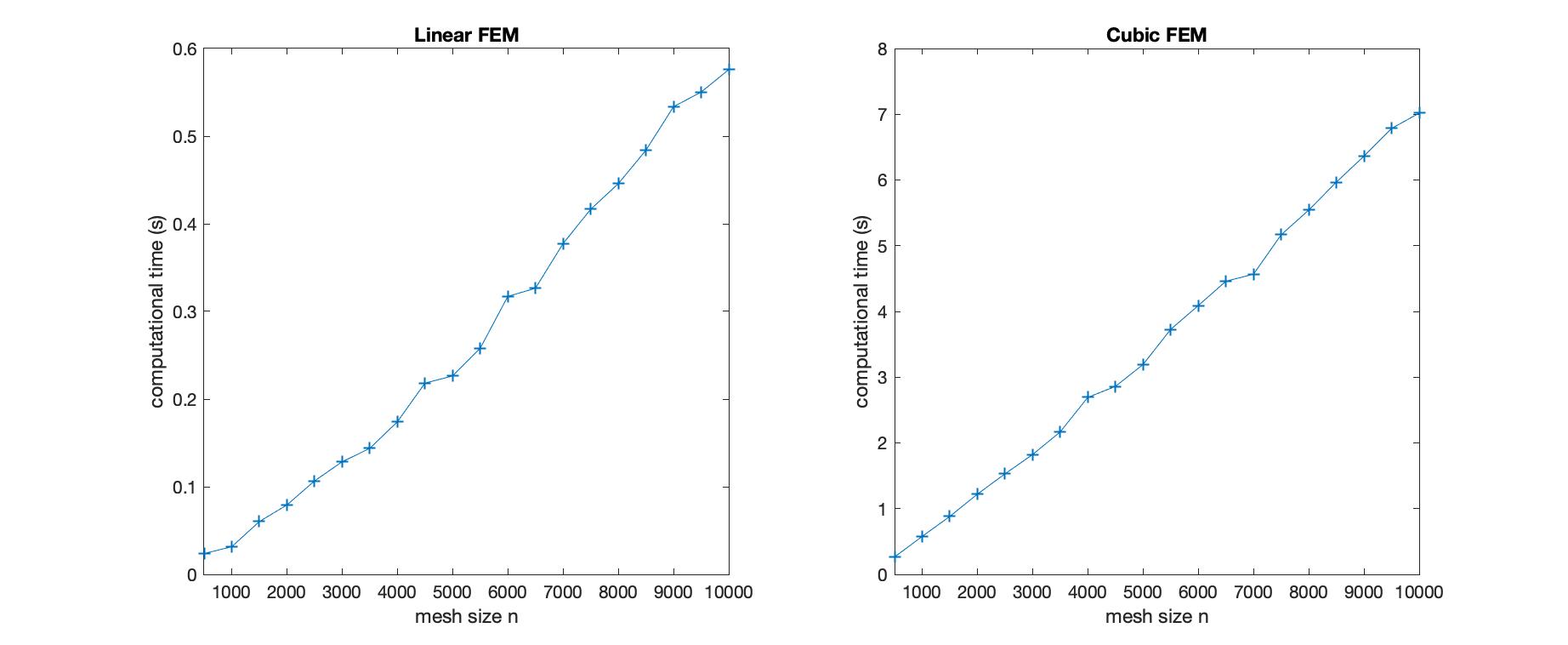}
	\caption{The computational time in seconds versus the mesh size $N$ for computing the first 50 LB eigenvalues of a unit sphere. Experiments run on a machine with 3 GHz Quad-Core Intel Core i5 and 8G RAM.}
	\label{complexity}
\end{figure}

Another immediate property drawn from (\ref{ABelements}) is that both matrices $\bm A$ and $\bm B$, being Gram matrices, are symmetric, which assures the estimated Laplace-Beltrami spectrum is real. There are computational benefits of solving the generalized eigenvalue problem (\ref{AB}) for symmetric and sparse matrices. For example, the Arnoldi algorithm  has a typical computational complexity of $\mathcal{O}(kN)$ to solve for the first $k$ eigenvalues, where $N$ is the matrix size \citep{ZhaoEDCTech}. This can also be seen from Figure \ref{complexity}, where the computational time for finding a fixed number of LB eigenvalues is linear in $N$, the mesh size, in both linear and cubic FEM methods. As expected, the cubic FEM takes longer as it works with larger matrices by adding additional nodes.

\subsubsection{Accuracy and convergence}
\label{accuracy}
\begin{figure}
	\centering
		\includegraphics[width=\textwidth]{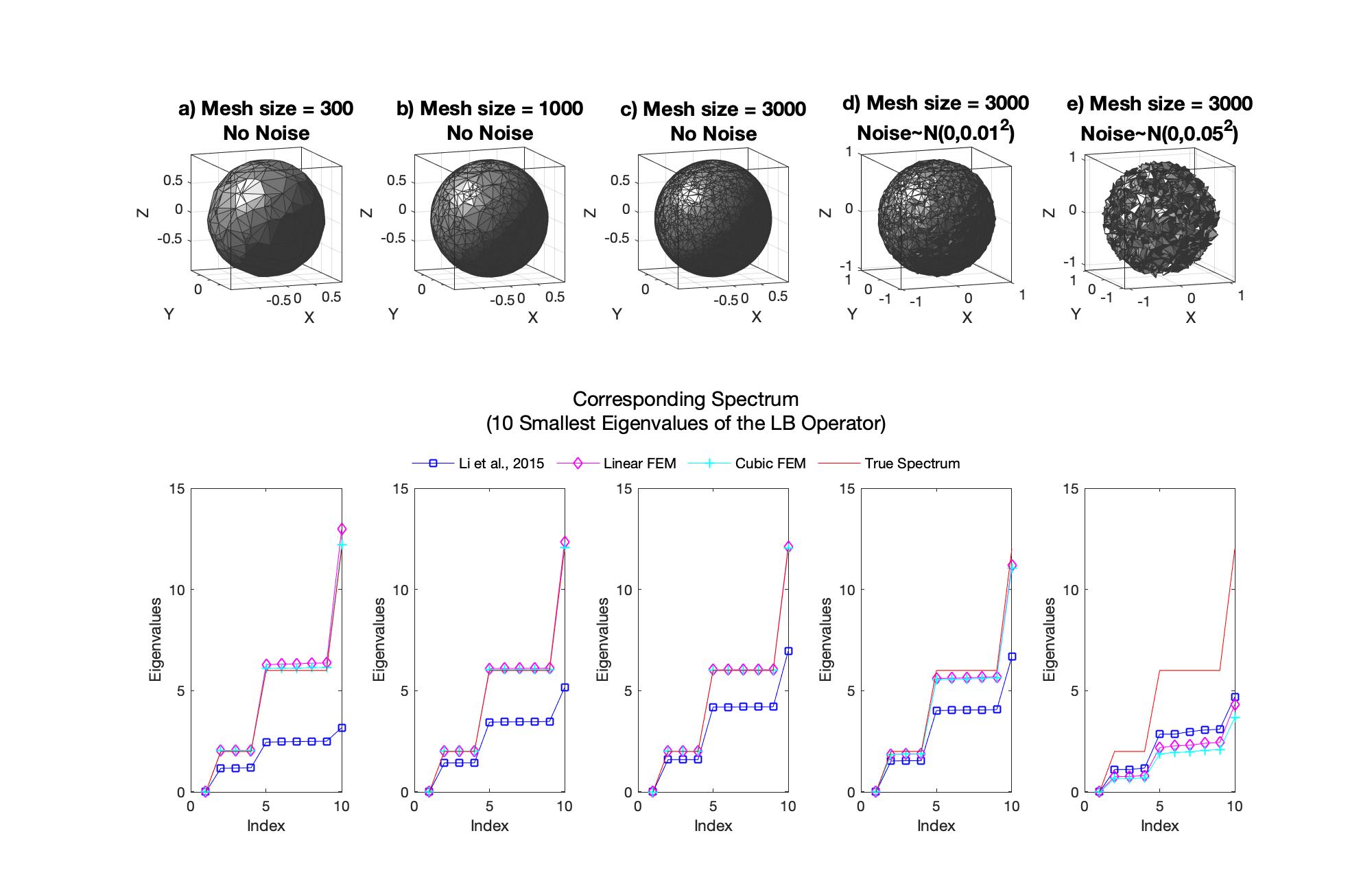}
	\caption{The leading spectrum of the FEM Laplacians versus that obtained from the heat kernel based Laplacian \citep{li2015localized} for a unit sphere under different conditions. The true (analytical) LB spectrum is also plotted for comparison. From left to right: a-c) are noise-free spheres with increasing mesh size. Though all estimated spectra converge to the true spectrum as the density of the mesh increases and approximates the manifold better, the FEM spectra are already very accurate with a mesh size as small as 300 points.  c-e) are the same sphere with increasing noise level, which show how all estimated spectra are affected by surface noise. A more accurately estimated LB spectrum will result in better detection properties when the spectrum is used for SPC or inspection purposes.}
	\label{FEMvsLocalized}
\end{figure}

\begin{figure}
	\centering
		\includegraphics[width=\textwidth]{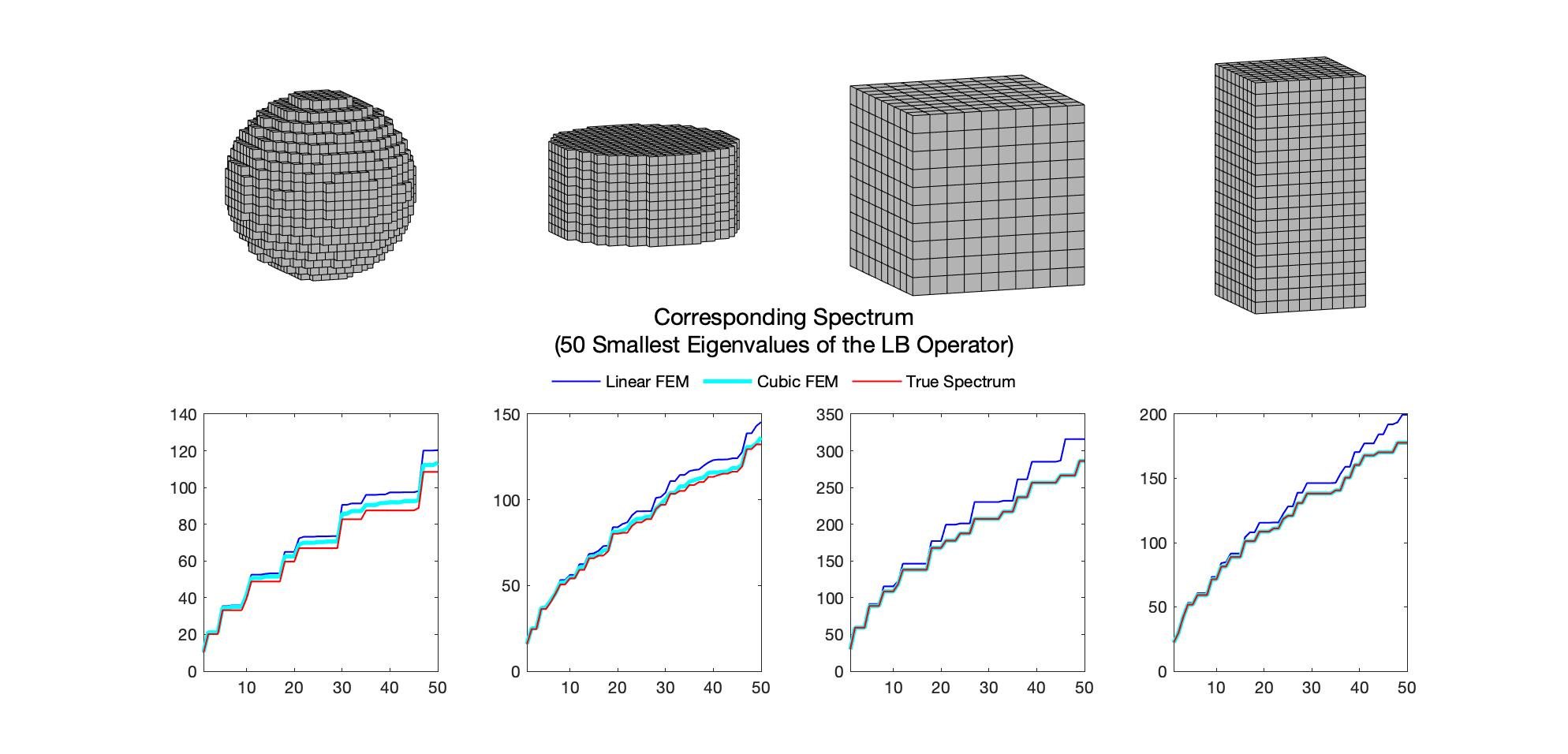}
	\caption{The first 50 LB eigenvalues of the FEM methods versus the analytical solutions (assuming the Dirichlet boundary condition). The solid ball has 10 voxels in the radius direction. The cylinder has 10 voxels in both the radius direction and height. The cube and the cuboid have $10\times10\times10$ and $10\times10\times20$ voxels, respectively. A thicker line is used for the cubic spectrum to avoid overlapping.}
	\label{FEMvol}
\end{figure}

\citet{reuter2006laplace} illustrate how the FEM methods appear to yield accurate results for the LB spectrum, with higher order shape functions resulting in more accurate estimations compared to the known analytic spectrum of some 3D objects. To verify this claim, we can compare the FEM LB spectrum with the methods used in \cite{ZhaoEDCTech} to compute the LB spectrum. Figure \ref{FEMvsLocalized} displays the first 10 eigenvalues in the  LB spectrum of a unit sphere (one of the few 3D objects for which the spectrum is known analytically) using different LB estimation methods. Note how much more accurate the FEM methods already are for a mesh size of only 300 points compared with the Localized Mesh Laplacian of \cite{li2015localized}. As it can be seen from the same figure, the FEM LB methods are robust to small surface noise and still sensitive to reflect surface changes caused by noise. Figure \ref{FEMvol} also shows how both linear and cubic  FEM LB spectra closely approximate  the analytical LB spectrum \citep{reuter2006laplacebook, helffer2016nodal} of several 3D objects in the voxel case. In the last two cases, since both the cube and cuboid can be represented exactly despite the small voxel numbers, the cubic FEM spectra are extremely close to the corresponding analytical LB spectrum. The difference between the first 50 LB eigenvalues obtained with the cubic FEM and the corresponding eigenvalues from the analytical LB operator has a norm of only 0.2391 for the cube and 0.0317 for the cuboid. A thicker line is used for the cubic spectrum so that it does not overlap with the red line representing the true spectrum.

\begin{figure}
	\centering
		\includegraphics[width=\textwidth]{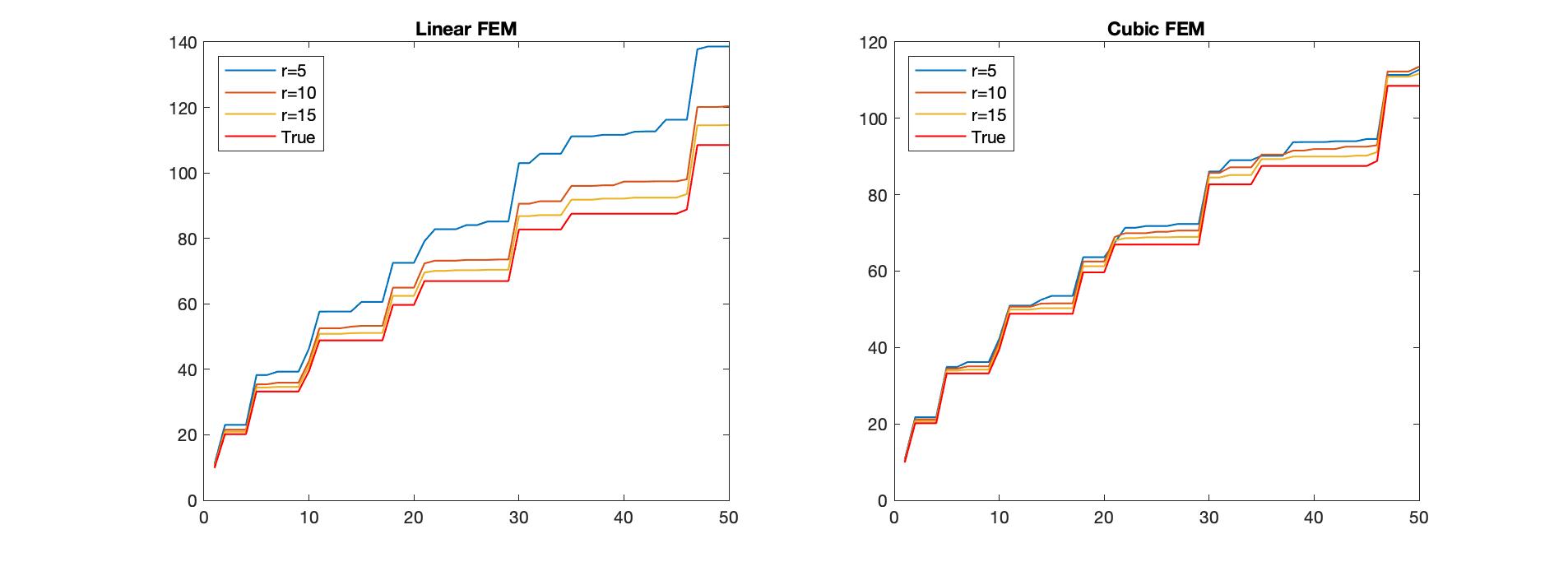}
	\caption{The first 50 eigenvalues of the FEM Laplacians converge to the true analytical LB spectrum (assuming the Dirichlet boundary condition) as we have more voxels in the radius direction, resulting in a more refined voxel approximation of the perfect solid ball with radius 1.}
	\label{FEM_converge}
\end{figure}

In practice, convergence of the FEM method is achieved either by refining the mesh, i.e., decreasing the mesh elements relative to the scanned object, or increasing the degree of the polynomial approximation.  \citet{reuter2006laplacebook} mentions how the convergence of the FEM method with shape functions of order $p$ behaves asymptotically with an error of order $\mathcal{O}(s^{p+1})$ as the largest mesh element size $s$ goes to zero (here, $s=\max{s_i}$ where $s_i$ is a measure of each element size, e.g., the length of the largest side or the radius of the largest inscribed circle in element $i$, see \cite{ihlenburg2006finite}). 
This fast convergence rate can be observed in Figure \ref{FEMvsLocalized} and Figure \ref{FEM_converge}, where both the linear and cubic FEM LB spectra are closer to the true spectrum as the mesh and the voxel representation become more refined. \cite{ihlenburg2006finite} emphasizes how convergence is due to the combined effect of the ``approximability'' of the shape functions and the numerical stability of the computations. The numerical stability is improved if the triangulation does not have wildly different sizes. The author provides different types of convergence theorems for FEM's applied to the solution of Helmholtz boundary problems which all hold under the condition that $\lambda s < 1$. Hence, determination of the eigenvalues in a relatively upper part of the spectrum (when sorted by magnitude) requires very large meshes with small and not wildly variable in size elements. Fortunately, the SPC spectral methods we develop utilize the lower, or leading, part of the spectrum only. 

\subsubsection{Computational and storage requirements}
With respect to the computational and storage costs for creating and storing the $\bm A$ and $\bm B$ matrices, the shape functions and their integrals can be calculated beforehand, as discussed at the end of section \ref{constructAB}. For the case of surface meshes,  we only need to calculate the metric tensor $\bf g$ for each triangle, which has the computational complexity of $\mathcal{O}(T)$, with $T$ being the number of triangles  in a mesh. To construct the $\bm A$ and $\bm B$ matrices given the metric tensors, the computational complexity is linear in the number of non-zero elements in $\bm A$ and $\bm B$, which is $3 \choose 2$ per triangle for linear FEM and $10 \choose 2$ per triangle for cubic FEM. Overall, the construction of the FEM Laplacians has order $\mathcal{O}(T)$ for both linear and cubic methods, with a larger coefficient for the cubic case. The constructions of the $\bm A$ and $\bm B$ matrices in the voxel case is much simpler than in the surface case, since the metric tensor $\bm g$ remains constant and does not need to be recalculated for each finite element. As a result, the construction of the $\bm A$ and $\bm B$ matrices only depends on the number of non-zero elements and has the computational complexity of $\mathcal{O}(v)$, with $v$ being the number of voxels denoted as ``active''. Again the cubic FEM is expected to have a larger coefficient for the computational complexity compared to the linear FEM due to the increased number of nodes per voxel as the matrices are less sparse. The computational cost of solving for the first eigenvalues is mentioned in section \ref{SparsitySymmetry}. 

The storage cost is similar to the computational cost of constructing the $\bm A$ and $\bm B$ matrices given the metric tensor $\bm g$, as it depends on the number of non-zero elements of each matrix. However, the storage cost is expected to be smaller because two adjacent triangles can share the same pair of points that counts as one non-zero element but calculated twice in the $\bm A$ and $\bm B$ matrices, once for each triangle they appear in. Take the linear FEM for a 2D closed mesh as an example, let $d_i$ be the degree of point $i$, which is the number of edges connected with point $i$ in a mesh, then row $i$ in matrix $\bm A$ or $\bm B$ has $(d_i+1)$ non-zero elements. Furthermore, since both $\bm A$ and $\bm B$ matrices are symmetric, only the lower or upper triangular part needs to be stored, so the smallest storage needed for matrix $\bm A$ or $\bm B$ is $\sum_i (d_i/2+1)$ in this particular case, which can be roughly seen as $\mathcal{O}(N)$, with $N$ being the mesh size. We want to point out that the linear FEM Laplacians are the most sparse that a discretized Laplacian could be, because only interactions between directly connected points are taken into account. The exact storage cost for cubic FEMs and 3D voxel cases is more complicated, but it is easy to see it will again be linear in the number of nodes, where the number of additional nodes for the cubic FEM depends on the number of edges (in both 2D and 3D cases) and the number of triangles (in the 2D case only). 

\subsubsection{Advantage of considering boundary conditions for the SPC of open meshes}
\begin{figure}
	\centering
		\includegraphics[width=\textwidth]{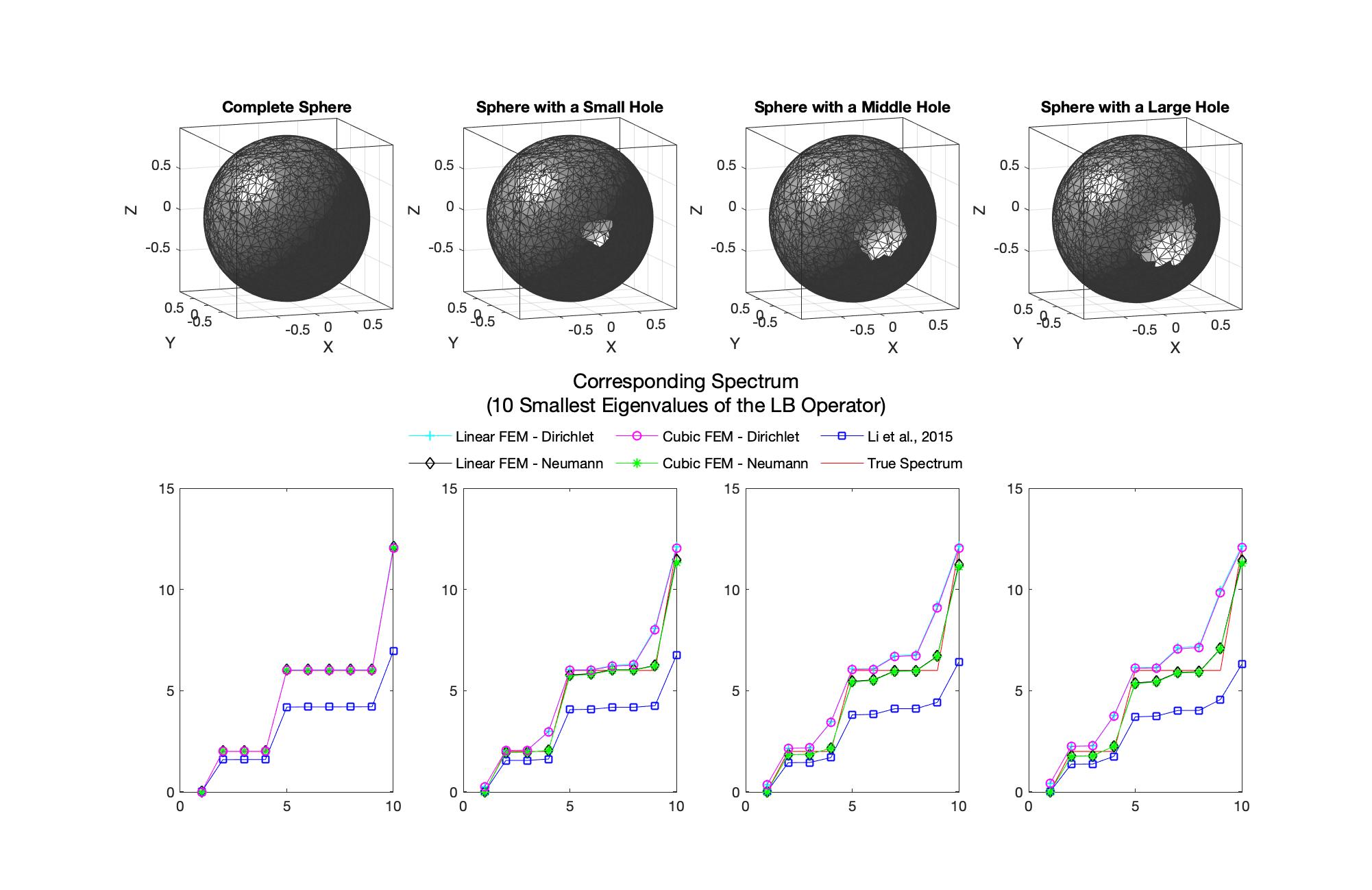}
	\caption{LB spectrum estimates for an open surface mesh with a hole. The graphs show the leading spectrum of the FEM Laplacians versus that of the localized Laplacian \citep{li2015localized} for a unit sphere with differently sized boundaries (holes). The FEM Laplacian spectra are closer to the true LB spectrum of a complete unit sphere than that of the localized Laplacian \citep{li2015localized}, especially under Neumann boundary conditions. In contrast, the Dirichlet FEM LB spectrum is more sensitive to the presence of the hole.}
	\label{boundary}
\end{figure}

Another advantage of the FEM methods over the methods used in ZD to estimate the LB spectrum is that the boundary conditions in the Helmholtz equation can be conveniently implemented, and this permits the control and inspection of parts from partial or open meshes that can easily result due to ``occlusion'' (unreachable areas to the scanner), in contrast to the spectral SPC method in \cite{ZhaoEDCTech} which can only be applied to closed objects with no boundaries. As discussed in Appendix A, the Dirichlet ($f\equiv 0$) and Neumann ($\frac{\partial f}{\partial n}\equiv 0$) boundary conditions simplify the weak form from (\ref{Green}) to (\ref{Green_simple}). Further, for the Dirichlet condition, we can just omit the boundary points and their corresponding shape functions, which are constantly zero over the whole surface and do not contribute to the $\bm A$ and $\bm B$ matrices at all. For the Neumann condition, \citet{reuter2006laplacebook} suggests simply treating the boundary points as inner points. Either boundary case allows for the statistical process monitoring of ``open'' meshes, as opposed to ``closed'' meshes using the FEM spectra. Figure \ref{boundary} plots the leading spectra of different LB approximations as well as the true LB spectra. The sequence of noise-free unit spheres have larger holes on their surface from left to right in the figure, and all estimated spectra are affected. The Dirichlet LB spectrum is more sensitive to the presence of the holes compared to the Neumann LB spectrum which changes less in the presence of the increasing holes, staying closer to the true (analytic) spectrum of the whole (hole-free) surface.  Under the same boundary condition, there is not much difference between the linear FEM and cubic FEM, since they both converge fast as shown in Figure \ref{FEMvsLocalized}. Both boundary conditions can be implemented for the voxel FEM methods, discussed above, in the same way. 

\section{Run length behavior}
\label{sec:4}
A standard performance metric of any SPC chart is the out of control {\em run length}, defined as the number of parts sampled between a defect (or out of control condition) occurs in a sequence of measured parts and when this is detected by the chart mechanism \citep{MontgomerySPC}. Also important is the in-control run length, defined as the number of parts sampled between false detections, when the process is actually in a state of control. One seeks short out of control run lengths and long in-control run lengths. Closed form expressions for the in and out of control run length distributions of most SPC charts are intractable, and it is customary to estimate the Average Run Length (ARL) and the standard deviation of the run length (SDRL) using Monte Carlo simulation. In this section we adopt the nonparametric control chart in \cite{chen2016distribution}, called the distribution-free multivariate exponentially-weighted moving average (``DFEWMA'') chart, and apply it to monitor changes in the first 15  eigenvalues of the estimated FEM LB spectra of parts via simulation. It is crucial to use a nonparametric chart, given that the non-normality (non-gaussianity) of the LB spectra of measured parts has been observed even in simulated surface meshes with isotropic, uncorrelated normal-distributed noise \citep{ZhaoEDCTech}. In Appendix B we provide a brief overview of the DFEWMA chart operation and its tuning parameters, which will be referred to in this section. To gain a more complete sense of the effectivity of the SPC chart using the estimated FEM spectra, we conduct the run length analysis based on simulated objects in different scenarios and compare the run length performance against some other previously proposed methods for SPC of 3D objects. We consider different practical cases, from varying the noise structure to generalizing the type of data, from 2D meshes to 3D voxels, in both cases including cases where the mesh is open. 

\subsection{Uncorrelated isotropic noise}
The simplest case is where the coordinate measurements contain noise, due to the combination of measurement error and manufacturing error, which is uncorrelated and isotropic in space. We show two examples under these conditions, a prototype part with three types of local defects as the out-of-control scenarios, and a series of cylindrical parts with a ``barrel-like'' shape controlled by $\delta>0$ as an out-of-control parameter. For the prototype part, we consider both ``closed'' meshes without boundaries and ``open'' meshes with boundaries. The following results show the FEM methods have outstanding performance in both cases.

\subsubsection{Prototype part}
The first example is a prototype part used in \citet{ZhaoEDCTech}, which is typical in an additive manufacturing process. Three types of defects are considered, namely two types of ``chipped'' corner parts and a part with a ``protrusion'' in one of the top edges, shown earlier in Figure \ref{ClosedParts}. To simulate manufacturing and measuring noise, isotropic $N(0.05^2\bf{I}^3)$ noise is added to the coordinate of each point and to simulate the case of unregistered meshes with {\em unequal} number of (non-corresponding) points, between 0 and 5 points are randomly deleted from each simulated part, resulting in mesh sizes of 1675-1680 points. Table \ref{ARL:ClosedPart} shows the average run lengths (ARL) and the standard deviation of the run lengths (SDRL) for the FEM methods compared to the \citet{li2015localized} Laplacian and a registration based method proposed by \citet{ZhaoEDCTech} as a benchmark. The in-control case uses a nominal ARL of 20 to avoid long simulation times, and it is achieved by all methods thanks to the DFEWMA control chart, which is easy to tune for a desired in-control ARL \citep{chen2016distribution}. Changing the DFEWMA chart design parameters such that the nominal in-control ARL is 200, most of the methods are able to signal within the first 5 defective parts when the process is out of control. Furthermore, both FEM methods are able to detect the defects quickly with minimal requirements on mesh sizes and mesh qualities, while the \citet{li2015localized} Laplacian needs a larger mesh with a higher mesh quality, obtained from the Loop subdivision algorithm (see \cite{loop1987smooth} and \cite{ZhaoEDCTech}) as a pre-processing step, to better capture the local defects. This indicates that the FEM methods are far more sensitive than the \citet{li2015localized} Laplacian to reflect local shape changes. 

\begin{table}
\begin{center}
\begin{tabular}{c|c|c|c|c}
\hline
 & In-control Part & Chipped \#1 & Chipped \#2 & Protrusion \\
\hline
Nominal In-control RL & 20 (19.49) & 200 (199.50) & 200 (199.50) & 200 (199.50)\\
Linear FEM (original mesh) & 20.16 (19.56) & 2.00 (0.03) & 2.00 (0.01) & 2.00 (0.00) \\
Cubic FEM (original mesh) & 20.12 (19.54) & 2.00 (0.04) & 2.00 (0.02) & 2.00 (0.00) \\
\citet{li2015localized} (original mesh) & 20.17 (19.89) & 158.12 (182.18) & 91.38 (135.31) & 3.65 (1.72)\\
\citet{li2015localized} (preprocessed mesh) & 20.49 (20.09) & 5.09 (2.77) & 4.44 (2.12) & 2.43 (0.51) \\
ICP &20.13 (19.20) & 2.00 (0.00) & 2.00 (0.00) &2.00 (0.00)\\
\hline
\end{tabular}
\caption{Run length performance of the DFEWMA SPC charts applied to the FEM LB spectrum for the test part examples with closed meshes. Results are obtained from 10,000 replications. Chart parameters were set at $m_0=100, w_{\text{\tiny min}}=1, w_{\text{\tiny max}}=10, \lambda=0.01,$ and $ \alpha=0.005$ for the out-of-control cases and $\alpha=0.05$ for the in-control case (resulting in shorter in-control run lengths to reduce simulation expense). First 15 LB operator eigenvalues were used. All methods achieve the nominal in-control run length, but there are notable differences in our of control performance, with the FEM methods and the use of the Iterated Closest Point (ICP) method (proposed in \cite{ZhaoEDCTech} as a benchmark) performing the best. The computational expense and non-convexity of the ICP formulation makes this method impractical. Due to the nature of the DFEWMA chart, a run length of 2.0 is the minimum possible that can be achieved.}
\label{ARL:ClosedPart}
\end{center}
\end{table}

\begin{figure}
	\centering
		\includegraphics[width=\textwidth]{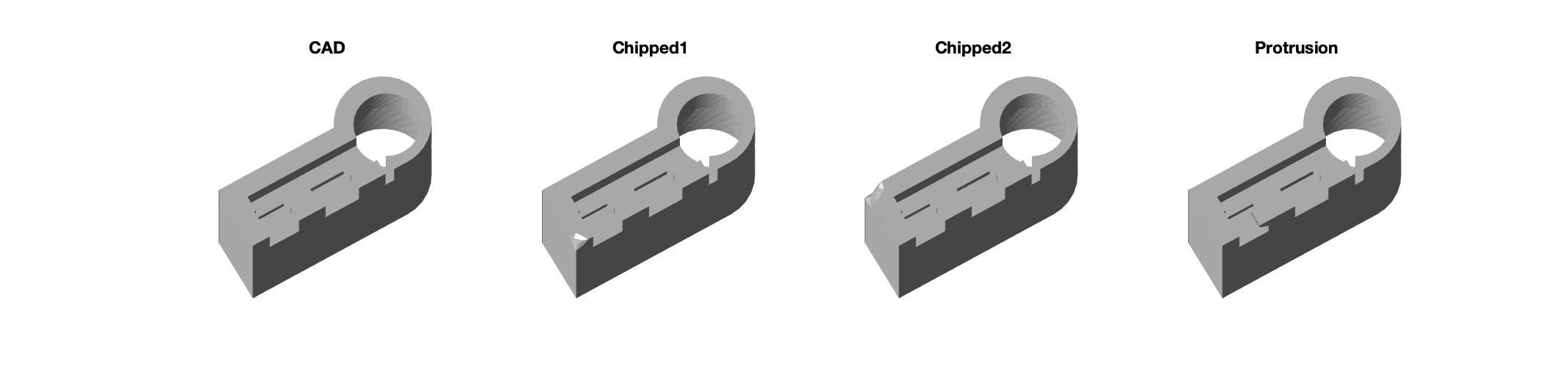}
	\caption{Prototype parts with open meshes used for the run length analysis in Table \ref{ARL:OpenPart}. The left most figure is the CAD model, while the other three figures show the different types of defects (chipped corners or a protrusion). All meshes are open and have boundaries, where the bottom of the part and a portion of the interior of the cylindrical region are missing, a situation common due to regions that are unreachable to a scanner.}
	\label{OpenParts}
\end{figure}

We also simulated cases with ``open'' meshes, as they show one of the greatest advantages of the FEM LB method. In this case, all the bottom of the mesh and a portion of the interior of the cylindrical region of the prototype part are deliberately omitted to better represent the practical case when a range scanner encounters unreachable regions of an object during a scan. The resulting meshes are shown in Figure \ref{OpenParts} and the corresponding estimated run length results using the FEM spectra are listed in Table \ref{ARL:OpenPart}. The Dirichlet boundary condition is applied because it is more sensitive to the ``holes'' in the mesh, as shown in Figure \ref{boundary}, and also results in smaller sparse matrices after deleting the rows and columns corresponding to the boundary points. From the table, both the linear FEM and cubic FEM are able to detect the out-of-control scenarios immediately, while achieving the nominal in-control run length behavior at the same time. By comparing the results with Table \ref{ARL:ClosedPart}, we can see open meshes have slightly longer out-of-control run lengths than closed meshes, but the difference is negligible. This case proved the applicability of the FEM methods for SPC of open meshes, a realistic case when a range sensor is mounted in a fixed position and cannot ``see'' the object from all perspectives.

\begin{table}
\begin{center}
\begin{tabular}{c|c|c|c|c}
\hline
 & In-control Part & Chipped \#1 & Chipped \#2 & Protrusion \\
\hline
Nominal In-control RL & 20 (19.49) & 200 (199.50) & 200 (199.50) & 200 (199.50)\\
Linear FEM & 20.61 (20.24) & 2.00 (0.07) & 2.00 (0.03) & 2.00 (0.05) \\
Cubic FEM & 20.15 (19.62) & 2.00 (0.06) & 2.00 (0.04) & 2.00 (0.05) \\
\hline
\end{tabular}
\caption{Run length performance of the DFEWMA charts applied to the FEM LB spectrum for the test part examples with open meshes. Results are obtained from 10,000 replications. Chart parameters were set at $m_0=100, w_{\text{\tiny min}}=1, w_{\text{\tiny max}}=10, \lambda=0.01,$ and $ \alpha=0.005$ for the out-of-control cases and $\alpha=0.05$ for the in-control case. First 15 LB operator eigenvalues were used.}
\label{ARL:OpenPart}
\end{center}
\end{table}

\subsubsection{``Barrel-like'' cylindrical parts}
To parameterize the out-of-control run length in a simple way, we consider cylindrical parts acquiring a more ``Barrel-like'' shape as an OC parameter $\delta>0$ increases. This shape is one of the typical out of control signals in a turning manufacturing operation of cylinders in a lathe process \citep{Colosimo2014}. To construct the out-of-control parts, using cylindrical coordinates we added a first harmonic with amplitude $\delta$ times the standard deviation of the noise to the radius, so the deformed radius at height $h$ becomes $10+0.05\delta\sin(h\pi/50)$, where 10 and 50 are the desired radius and height of the cylindrical parts, respectively. Isotropic white noise $N(0, 0.05^2 {\bf I}_3)$ is added to the coordinates of the points. Table \ref{ARL:OC-cylinder} compares the ARL and SDRL of the different methods, including the various LB spectra and the ICP based method, as a function of the OC parameter $\delta$. The FEM spectra detect small changes into the ``barrel'' shape of the cylinder (small $\delta$) faster than the  \citet{li2015localized} spectrum. The ICP method is less efficient because it only considers the average deviation from the CAD model, and small differences can be easily masked by the overall natural variability of the in-control process. 

\begin{table}[H]
\begin{center}
\begin{tabular}{c|c|c|c|c|c|c|c|c}
\hline
\multirow{2}{*}{} & \multicolumn{2}{c|}{Linear FEM} & \multicolumn{2}{c|}{Cubic FEM} & \multicolumn{2}{c|}{\citet{li2015localized}} & \multicolumn{2}{c}{ICP objective} \\
\cline{2-9}
	& ARL & SDRL & ARL & SDRL & ARL & SDRL & ARL & SDRL  \\
\hline
$\delta=0$ & 20.07 & 19.57 & 20.09 & 19.60 & 20.46 & 20.24 & 20.25& 19.88 \\
\hline	
$\delta=0.0005$ & 5.67 & 2.70 & 4.03 & 1.76 & 10.79 & 9.94 & 83.21 & 122.65 \\
\hline	
$\delta=0.005$ & 2.03 & 0.16 & 2.00 & 0.00 & 2.03 & 0.19 & 39.76 & 65.89 \\
\hline
$\delta=0.5$ & 2.00 & 0.00 & 2.00 & 0.00 & 2.00 & 0.00 & 31.49 & 51.87 \\
\hline
$\delta=1$ & 2.00 & 0.00 & 2.00 & 0.00 & 2.00 & 0.00 & 5.19 & 3.01 \\
\hline
$\delta=2$ & 2.00 & 0.00 & 2.00 & 0.00 & 2.00 & 0.00 & 2.03 & 0.17 \\
\hline
$\delta=3$ & 2.00 & 0.00 & 2.00 & 0.00 & 2.00 & 0.00 & 2.00 & 0.00 \\
\hline
$\delta=10$ & 2.00 & 0.00 & 2.00 & 0.00 & 2.00 & 0.00 & 2.00 & 0.00 \\
\hline
\end{tabular}
\caption{Phase II out-of-control run length performance of the DFEWMA charts applied to the different LB spectra and ICP objective for barrel-shaped cylindrical parts. 10,000 replications, each with 100 IC cylinders followed by a sequence of defective cylinders until detection.  The DFEWMA chart parameters  are: $m_0=100, w_{\text{\scriptsize min}}=1, w_{\text{\scriptsize max}}=10$, and $\lambda=0.01$. $\alpha=0.005$ for $\delta>0$ and $\alpha=0.05$ for $\delta=0$, corresponding to an in-control ARL of 200 and 20, respectively. First 15 LB operator eigenvalues were used, and mesh sizes varied between 1995 and 2005 non-corresponding points.}
\label{ARL:OC-cylinder}
\end{center}
\end{table}

\subsection{Spatially correlated nonisotropic noise}
\label{correlatedRL}
A more general case in practice is when the noise is correlated and non-isotropic, created, e.g., by manufacturing noise that changes spatially on the surface of the objects depending on how the cutting tool operates. To analyze this case, we repeated the analysis for the barrel-shape cylinders to show the run length performance of the FEM methods under correlated spatial noise. The defective shape is introduced in the same way as before, so the deformed radius at height $h$ is still $10+0.05\delta sin(h\pi/50)$, where $10$, $50$, and $0.05$ are the nominal radius, nominal height, and the standard deviation of noise, respectively. At each point $\boldsymbol{p}_i=\begin{pmatrix} p_{i,x}, p_{i,y}, p_{i,z}\end{pmatrix}$, non-isotropic and spatially correlated noise $\boldsymbol{e}_i=\begin{pmatrix} e_{i,x}, e_{i,y}, e_{i,z}\end{pmatrix}$ is added to the point coordinate. The covariance functions between different noise terms are:
$$Cov(e_{i,k}, e_{j,l})=\begin{cases} \sigma_1^2e^{-|p_{i,k}-p_{j,l}|/r_k} & \text{ if } i\neq j, k=l \\
\sigma_1^2+\sigma_2^2 & \text{ if } i=j, k=l \\
0 & \text{ if } k\neq l
\end{cases}$$
Here $i, j$ are point indices and $k, l\in\{x,y,z\}$ indicate the axes. To keep a constant total level of noise, we fixed $\sigma_1^2+\sigma_2^2=0.05^2$. Similarly to the previous example, the mesh sizes for the cylindrical parts randomly vary between 1995 and 2005 points to model non-corresponding, different size meshes, and the first 15 eigenvalues are used. Table \ref{ARL:OC-cylinder-nonisotropic} shows the results. As it can be seen, again the FEM spectra have a similar and slightly better detection compared to the \citet{li2015localized} spectrum when $\sigma_1=0.02$, or in other words, $\sigma_2\neq 0$. When $\sigma_1=0.05$, the FEM spectra are affected by the spatial correlation and fail to detect small changes quickly. This is likely due to the irregular triangulations that result from the correlated data, as irregular meshes cause stability problems for FEM. Applying a mesh pre-processing method prior to computation of the LB spectra notably improves the run length performance.


\begin{table}[H]
\begin{center}
\begin{tabular}{c|c|c|c|c|c|c|c|c|c}
\hline
& \multicolumn{3}{c|}{$\sigma_1^2=0.02^2, r_x=r_y=2.6$} & \multicolumn{3}{c|}{$\sigma_1^2=0.02^2, r_x=r_y=5.2$} & \multicolumn{3}{c}{$\sigma_1^2=0.05^2, r_x=r_y=2.6$} \\
\cline{2-10}
\multicolumn{1}{c|}{$\delta$} & 0.0005 & 0.005 & 1 & 0.0005 & 0.005 & 1 & 0.0005 & 0.005 & 1 \\
\hline
Linear FEM & 5.41 & 2.10 & 2.00 & 5.40 & 2.11 & 2.00 & 67.96 & 3.71 & 2.00\\
(original) & (2.54) & (0.30) & (0.00) & (2.46) & (0.31) & (0.00) & (107.57) & 0.65 & (0.00)\\
\hline
Linear FEM & 6.27 & 2.00 & 2.00 & 6.35 & 2.00 & 2.00 &16.44 & 2.06 & 2.00\\
(preprocessed) & (3.75) & (0.01) & (0.00) & (3.96) & (0.00) & (0.00) & (23.04) & (0.24) & (0.00)\\
\hline
Cubic FEM & 3.97 & 2.00 & 2.00 & 3.96 & 2.00 & 2.00 & 111.46 & 14.13 & 2.46\\
(original) & (1.69) & (0.00) & (0.00) & (1.70) & (0.00) & (0.00) & (150.76) & (25.36) & (0.50)\\
\hline
Cubic FEM & 5.21 & 2.00 & 2.00 & 5.13 & 2.00 & 2.00 & 5.90 & 2.00 & 2.00\\
(preprocessed) & (2.66) & (0.00) & (0.00) & (2.59) & (0.00) & (0.00) & (3.16) & (0.00) & (0.00)\\
\hline
\citet{li2015localized} & 11.50 & 2.03 & 2.00 & 9.57 & 2.02 & 2.00 & 8.48 & 2.01 & 2.00 \\
(original) & (13.89) & (0.19) & (0.00) & (8.19) & (0.14) & (0.00) & (6.85) & (0.10) & (0.00) \\
\hline 
ICP objective & 198.08 & 204.63 & 14.54 & 199.92 & 201.17 & 15.62 & 202.29 & 203.91 & 26.61 \\
(original) & (196.86) & (200.75) & (17.66) & (197.25) & (199.72) & (19.37) & (201.40) & (199.70) & (38.84) \\
\hline 
\end{tabular}
\caption{Phase II out-of-control run length performance of the DFEWMA charts applied to the different LB spectra and ICP objective with barrel-shaped cylindrical parts under spatially correlated, non-isotropic noise. 10,000 replications, each consisting of 100 IC cylinders followed by defective cylinders until detection. DFEWMA charts were applied to both LB spectrum method and ICP with parameters: $m_0=100, w_{\text{\scriptsize min}}=1, w_{\text{\scriptsize max}}=10, \lambda=0.01$ and $\alpha=0.005$, corresponding  to an in-control ARL of 200. First 15 LB operator eigenvalues used, mesh size varied between 1995 and 2005 points. {$r_z=16.7$ for all three cases.}}
\label{ARL:OC-cylinder-nonisotropic}
\end{center}
\end{table}

\subsection{SPC performance comparisons against registration based SPC methods}
We also compare the Phase II run length behavior of our FEM LB spectrum methods with an existing SPC method for 3D geometrical data due to \cite{Colosimo2014}, which is based on Gaussian Processes. It should be pointed out that this is a method aimed at contact sensed data and hence assumes small, equally sized meshes with corresponding points from part to part distributed in a lattice pattern, and is a method that performs GPA registration of the points first. Their  method cannot handle the harder problem of non-contact data, where the numbers of points per part varies and points do not correspond from part to part, and would have trouble if points did not form a lattice. Still, Table \ref{tab:9} shows how that the spectral FEM SPC method is very competitive in these unfavorable circumstances, and even sometimes provides better run length performance. The FEM methods once again show better run length performance in this case than the localized Laplacian spectral method used by \cite{ZhaoEDCTech}.

\begin{table}[H]
\vspace{-0.1cm}
\begin{center}
\begin{tabular}{c|c|c|c|c|c|c}
\hline
\multicolumn{2}{c|}{\multirow{2}{*}{}} & \multicolumn{3}{c|}{LB Spectra} & \multirow{2}{*}{GP$_\text{sub\_unif}$} & \multirow{2}{*}{GP$_\text{sub\_lh}$} \\
\cline{3-5}
\multicolumn{2}{c|}{}  & Linear FEM & Cubic FEM & \citet{li2015localized} & & \\
\hline
\multirow{2}{*}{In Control} & ARL & 100.57 & 96.63 & 99.85 & 99.69  & 100.77  \\
 &(SDRL) & (100.71) & (91.74) & (94.33) &  (97.41)& (100.94)  \\
\hline
Quadrilobe & ARL & 4.17 & 3.60 & 6.29 &4.70  & 1.39  \\
 $\delta=0.00185$ &(SDRL) &(1.20) & (1.04) & (2.65) &  (3.97)& (0.77)  \\
\hline
Half frequency & ARL & 1.85 & 1.87 & 3.27 & 14.11  & 4.51 \\
 $\delta=0.00075$ &(SDRL) &(0.51) & (0.53) & (1.44) &  (13.44)& (4.05)  \\
\hline
\end{tabular}
\caption{Out-of-control run length performance comparisons of different LB spectra versus two Gaussian processes (GPs) models studied in \cite{Colosimo2014}. Results obtained from 1,000 replications. DFEWMA chart parameters are: $m_0=100, w_{\text{\scriptsize min}}=1, w_{\text{\scriptsize max}}=10, \lambda=0.01$ and $\alpha=0.01$, corresponding to an in-control ARL of 100. First 15 LB operator eigenvalues were used, and the mesh size is fixed to 1054 points. {Mesh pre-processing based on the Loop method is applied for all three LB spectrum methods, resulting in around 2000 points.} Results of the GP methods were originally reported in Table 3, in \cite{Colosimo2014}. }
\label{tab:9}
\end{center}
\end{table}

\subsection{Voxel data}
\begin{figure}
	\centering
		\includegraphics[width=\textwidth]{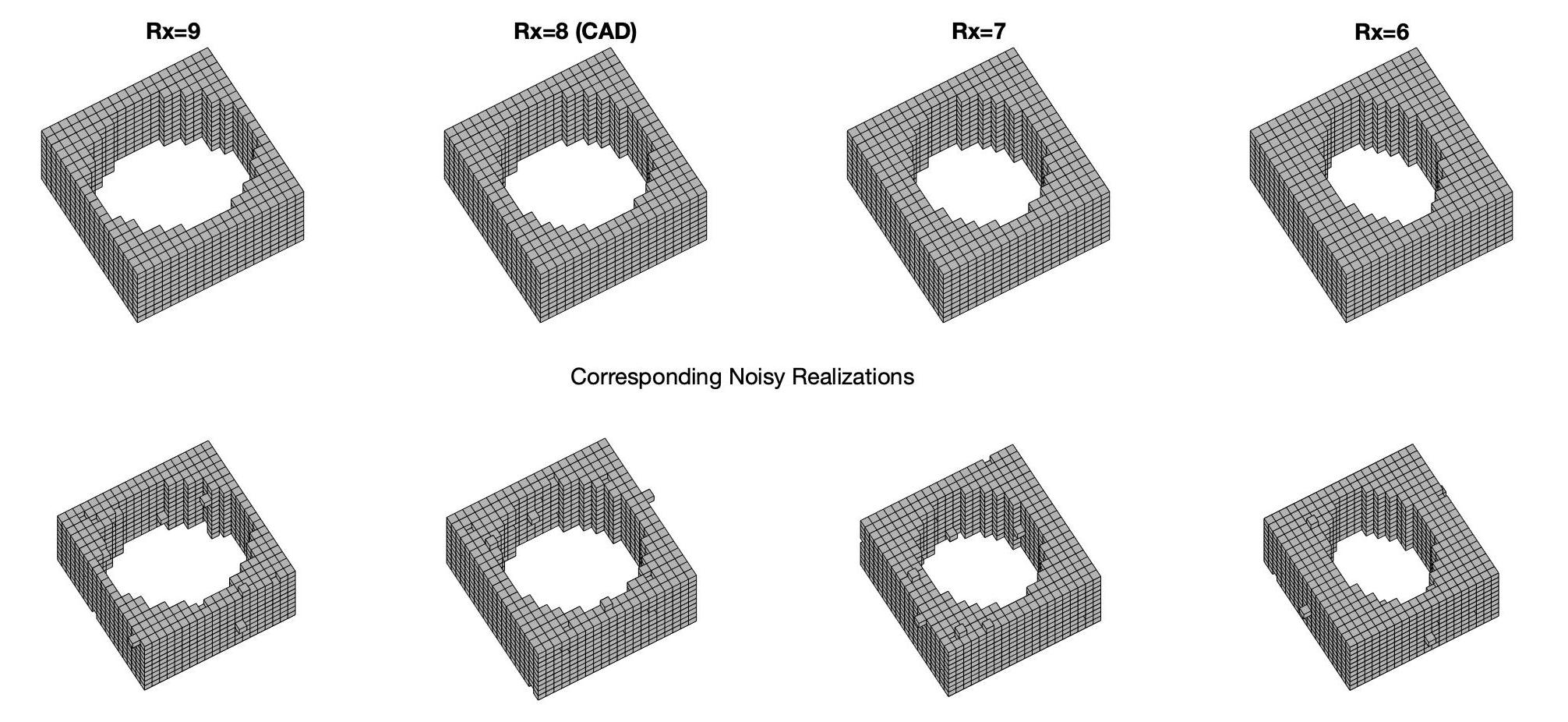}
	\caption{Voxel objects used for the SPC run length analysis in Table \ref{ARL:Voxel}. The first row shows the noise-free models of four parts with an inner cylinder of different eccentricity and orientation.   The second row of figures shows sample realizations of the corresponding parts when noise is added. The CAD model is the second part from the left, with a perfect circular cylinder (eccentricity=0) that corresponds to a radius value $Rx=8$.  The other three parts have elliptical cross-sections of different eccentricities of 0.4581, 0.4841, and 0.6614, corresponding to $Rx$ parameter values of 9, 7 and 6, respectively.}
	\label{Voxel_RL}
\end{figure}

Finally we consider a run length analysis for 3D voxel data. To the best of our knowledge, there has not been a Statistical Process Control scheme proposed for voxel sensor data in the literature, so only the performance of the linear and cubic FEM spectra are compared. To simulate the volumetric datasets obtained via CT scans of a part with inner features, we consider a cube with a hollow cylinder inside it, see Figure \ref{Voxel_RL}.  The dimension of the cube is $20\times20\times10$ voxels, and the nominal radius of the hollow cylinder is 8 voxels, shown in the second column. For the out-of-control scenarios, we vary the radius of the cylinder along one of the axis, denoted by $Rx$, to make increasingly more elliptic cylinders. The in-control part has a cylindrical hole with a cross section with eccentricity$ =0$ (parametrized with a value of $Rx=8$). We chose $Rx=9$, $Rx=7$, $Rx=6$ as three types of defectives, corresponding to ellipses of different orientations and eccentricities, namely, 0.4581, 0.4841, and 0.6614, respectively (recall circles have eccentricity $=0$ and for parabolas eccentricity $=1$, with ellipses in the range $0 <$ eccentricity $<1$.).

For the volumetric run length simulations, noise was added differently than from the 2D meshes. CT scans usually return four values for each volume element, the $x y z$ coordinates and an additional value giving the opacity or intensity of the material. In our case, we will assume a rendering algorithm is applied such that each voxel is classified simply as ``active'' or ``inactive'' based on its intensity compared against a given threshold value. If the intensity is higher, then it indicates that the voxel contains enough material in it and should be denoted as ``active''. Therefore, noise is more likely to occur in the voxels near the boundary of the object being scanned due to the natural variability of the intensity measurement. For this reason, we added noise in two different ways, making ``active'' boundary voxels that in reality should be ``inactive'', and vice versa. The level of noise is parametrized by the maximum number of voxels allowed to have their ``active/inactive'' statuses switched, specified in column ``Max Noise'' of Table \ref{ARL:Voxel}. For example, when ``Max Noise'' is 25, then for each noisy realization, an integer between 1 and 25 is randomly selected with equal probability and that particular number of ``active'' boundary voxels are randomly selected to be ``inactive''. Next, this procedure is repeated to select ``inactive'' boundary voxels to be ``active''. Simulated parts with noise are shown in the bottom row of plots in Figure \ref{Voxel_RL}.

The run length results of applying the DFEWMA nonparametric chart for 3D solids with different levels of noise and types of defects are summarized in Table \ref{ARL:Voxel}. The Dirichlet boundary condition is applied because it gives smaller Laplacian matrices, which is beneficial especially in the voxel case since the number of vertices increases dramatically as we have more voxels. Overall, the FEM spectra have excellent detection power, and the cubic FEM is consistently outperforming the linear FEM because it is more accurate as stated in Section \ref{accuracy}. As expected,  increasing noise level makes it harder and results in slightly longer ARLs and SDRLs. Among the 3 types of defects, $Rx=9$ is the hardest to detect, because it has the smallest eccentricity and hence is the closest to the CAD model. On the other hand, $Rx=6$ with eccentricity 0.6614 can be easily detected by the 3D FEM spectra regardless of the noise level.
\begin{table}[H]
\begin{center}
\begin{tabular}{c|c|c|c|c|c}
\hline
\multirow{2}{*}{Max Noise} & \multirow{2}{*}{$Rx$} & \multicolumn{2}{c|}{Linear FEM} & \multicolumn{2}{c}{Cubic FEM} \\
\cline{3-6}
&	& ARL & SDRL & ARL & SDRL  \\
\hline
\multirow{3}{*}{25}  & 9 & 2.00 & 0.02 & 2.00 & 0.00 \\
\cline{2-6}	
 & 7 & 2.00 & 0.00 & 2.00 & 0.00  \\
 \cline{2-6}	
 & 6 & 2.00 & 0.00 & 2.00 & 0.00  \\
\hline
\multirow{3}{*}{50}  & 9 & 2.53 & 0.69 & 2.23 & 0.44  \\
\cline{2-6}	
 & 7 & 2.00 & 2.00 & 2.00 & 0.00 \\
 \cline{2-6}	
 & 6 & 2.00 & 0.00 & 2.00 & 0.00   \\
\hline
 \multirow{3}{*}{100}  & 9 & 4.67 & 2.52 & 3.95 & 1.89  \\
\cline{2-6}	
 & 7 & 2.03 & 0.16 & 2.00 & 0.03 \\
 \cline{2-6}	
 & 6 & 2.00 & 0.00 & 2.00 & 0.00  \\
\hline
\end{tabular}
\caption{Phase II out-of-control run length performance of the DFEWMA charts applied to the linear and cubic FEM LB spectra with voxel data. The Dirichlet boundary condition is applied. 10,000 replications, each consisting of 100 IC parts followed by defective parts until detection. DFEWMA charts were applied with parameters: $m_0=100, w_{\text{\scriptsize min}}=1, w_{\text{\scriptsize max}}=10, \lambda=0.01$ and $\alpha=0.005$, corresponding to an in-control ARL of 200. First 15 LB operator eigenvalues used, the number of voxels varied around 1600 to 2600.}
\label{ARL:Voxel}
\end{center}
\end{table}

\section{Conclusions}
\label{sec:5}
We have presented a new approach for the Statistical Process Control of 3-dimensional parts whose metrology is acquired with either range sensors (surface data) or CT scanners (volumetric data). The new approach is based on the computation of the Laplace-Beltrami operator spectrum, an operator that codifies the geometrical properties of an object around a point. The LB spectrum, being an intrinsic geometrical property of an object, permits the comparison of different parts without the need of ``part localization'' (registration) algorithms, which are hard nonconvex optimization problems whose ad-hoc solutions may result in increased noise \citep{ZhaoEDCTech}. In contrast to the spectral method recently proposed by \cite{ZhaoEDCTech}, the new method, based on estimating the LB spectrum from solving a Helmholtz boundary equation via FEM, is more accurate, can be estimated to do SPC on both mesh and voxel data, and can be applied to the important practical case of ``open'' meshes with holes, thanks to the explicit incorporation of the boundary conditions in the PDE. We have shown how the SPC run length performance of a nonparametric control chart that uses the FEM LB spectrum as a ``profile'' feature from part to part, is very competitive and in most cases, better, than that of existing state of the art SPC methods for 3D data (without requiring registration/part localization). The method was also demonstrated on meshes with boundaries, typical when the scanner is mounted at a fixed position and does not have reach to all the part, and on 3D volumetric data, returned by CT scans. The cubic FEM method provides consistently better run length performance than the linear FEM, at a higher storage and computational cost. Fortunately, the linear FEM has almost as good  run length performance as the cubic FEM approach.

As can be seen from our discussion on computational and storage requirements (section \ref{accuracy}), both the computational and storage cost mainly depend on the number of nodal points, which directly affects the dimension of the $\bm A$ and $\bm B$ matrices. This can potentially be a limitation for the cubic FEM method, especially for the 3D volumetric case, where two additional nodes are inserted for each edge. The second limitation for the FEM methods is revealed in section \ref{correlatedRL}, where they are less sensitive to detect part defects under stronger correlated noise, although the effect of this problem is reduced when a pre-processing of the mesh is applied prior to our methods. Still, this reflects the  dependency of FEM methods on the properties of the mesh, well-known in the field of PDEs, which, in our case, are reflected in the ability of a SPC chart mechanism to detect an out of control state.

One potential limitation of the spectral FEM methods in general, not directly related to our SPC proposal, is that they return two sparse matrices ($\bm A$ and $\bm B$) instead of one discretized Laplacian matrix $\bm L$, which, though does not affect the eigenvalues and hence our methods, may impose difficulties if one wishes to derive other geometrical properties from the estimated Laplacians. For example, equation (\ref{LBnormal}) shows how the mean curvature at a point can be estimated by simply multiplying the LB operator times the point coordinates (since $\Delta_\mathcal{M}{\bf p}(u,v) = 2H{\bf n}(u,v)$). However, this cannot be done for the FEM Laplacians since matrix $\bm B$ is not invertible and hence $\bm L = \bm B^{-1} \bm A$ is not possible to compute.

\subsection*{Supplementary materials}

MATLAB code that implements the FEM LB spectrum estimation methods for both mesh and voxel cases is provided, including the models used in this paper.\\

\bibliographystyle{model5-names}

\bibliography{refs}

\section*{Appendix A . Some details on the variational solution of Helmholtz equation}

To get the weak or variational form of Helmholtz equation (\ref{Helmholtz}), we first multiply it times a test function $\phi\in C^2$
\begin{equation}
\phi\Delta f=\lambda\phi f
\end{equation}
Integrating both sides of the equation over the surface area and applying Green's first identity (a direct consequence of Gauss' divergence theorem, see \cite{Marsden}, p. 475) yields:
\begin{equation}
\label{Green}
\int_\mathcal{M}\phi\Delta f dV=\int_{\Gamma}\phi~(\nabla f\cdot \nn)\; ds-\int_\mathcal{M} \nabla \phi \cdot \nabla f\; dV=\int_\mathcal{M}\lambda\phi f dV
\end{equation}
where $dV$ is either the surface element on a 2-dimensional manifold (surface) or a volume element in a 3-manifold $\mathcal{M}$, $\Gamma$ is the boundary of $\mathcal{M}$, $ds$ is either a length or area element on the boundary $\Gamma$ and $\nabla f\cdot \nn=\frac{\partial f}{\partial \nn}$. Both the Dirichlet ($f, \phi\equiv 0$) and Neumann ($\frac{\partial f}{\partial n}\equiv 0$) boundary conditions satisfy
$$\int_{\Gamma}\phi~(\nabla f\cdot \nn)\; ds=0$$
which simplifies (\ref{Green}) to the so-called Garlekin, weak, or variational form of the PDE, equation (\ref{Green_simple}). 

The underlying variational problem solved by the weak form equation (\ref{Green_simple}) is not that frequently discussed in applied FEM references. \cite{Strang} indicate how the equation of the weak form corresponds to the Euler-Lagrange equation of a minimum energy functional, which, for our Helmholtz problem $(\Delta - \lambda)u \equiv {\mc L} u =0$ is of the form:
\[
I(u) = \la {\mc L}u, u \ra  = \int_{\mc M}  {\mc L}u\; u \; dV
\]
A perturbation around the minimizer $f$ of $I(u)$ is next introduced of the form $f+ \epsilon \phi$ where $\epsilon$ is arbitrarily small. It can then be shown that the variation of the functional $I(u)_1$ is minimized when
$
\la {\mc L}f, \phi \ra = 0,
$
but note how this expression yields precisely the weak formulation, equating (\ref{Green}) to zero. 

\section*{Appendix B. Operation and parameters of the DFEWMA SPC chart}

All the run length analysis in the paper were obtained with the  \cite{chen2016distribution} distribution-free multivariate exponentially-weighted moving average (``DFEWMA'') chart. This is a chart that operates in the on-line or ``Phase II'' stage typical of SPC methods. We refer readers to \cite{ZhaoEDCTech} for methods for  ``Phase I'', or the parameter learning stage with in-control data. The Phase I method shown there can be directly applied with the more accurate and flexible FEM LB spectra presented in this paper, and was not discussed in the present paper for conciseness.  

The DFEWMA chart is a distribution-free multivariate control chart. Suppose $m_0$ acceptable parts are available from Phase I, and we want to test the $n$th manufactured part in Phase II assuming all previous $n-1$ parts do not trigger any alarm. Let the first $p$ estimated LB eigenvalues of each part $i$ be ${\bf X_i}\in\mathbb{R}^p$, then all existing parts can be represented by $\bf X_1, \cdots, \bf X_{m_0}, \bf X_{m_0+1}, \cdots, \bf X_{m_0+n}$. The chart calculates the following statistic 
$$
T_{jn}(w,\lambda)=\frac{\sum_{i=n-w+1}^n (1-\lambda)^{n-i} R_{jni}-\mbox{E}\left[\sum_{i=n-w+1}^n (1-\lambda)^{n-i} R_{jni}\right]}{\sqrt{\mbox{Var}\left(\sum_{i=n-w+1}^n (1-\lambda)^{n-i} R_{jni}\right)}}
\label{DFEWMA}
$$
where $R_{jni}$ is the rank of the $j$th eigenvalue from the $i$th part, $ X_{i,j}$, among the $j$th eigenvalues of all parts ranging from $X_{1,j}$ to $X_{m_0+n,j}$. Here $w$ is the window size and $\lambda$ is the weight in this EWMA-type of chart. In our run length analyses, we chose $\lambda=0.01$, and $w=\min\{\max\{n,w_{\min}\}, w_{\max}\}$ with $w_{\min}=1$, $w_{\max}=10$ for quicker detection of small changes. The use of a window implies that the quickest possible detection is greater than one part or sample. Both the expectation and variance terms in $T_{jn}(w,\lambda)$ can be analytically derived, see more details in \cite{ZhaoEDCTech}. The DFEWMA chart monitors the sum of squares $T_n(w,\lambda)=\sum_{j=1}^p T_{jn}^2(w,\lambda)$ given that differences in {\em all} the first $p$ eigenvalues should be considered jointly.
\end{document}